\documentclass[english,aps,prb,superscriptaddress,reprint,longbibliograhy]{revtex4-2} 

\usepackage{ulem}
\usepackage{graphicx}
\usepackage{hyperref}
\usepackage{amssymb}
\usepackage{dcolumn}
\usepackage{float}
\usepackage{booktabs}
\usepackage{amsmath} % or simply amstext
\usepackage[utf8]{inputenc}
\usepackage{lmodern}
\usepackage{makecell}
\usepackage{marginnote}

\usepackage{natbib}
\usepackage{float}
\usepackage{placeins}
\newcommand{\e}{\mathrm{e}}

\DeclareMathAlphabet{\bi}{OML}{cmm}{b}{it}

\def\ba{\begin{aligned}}
	\def\ea{\end{aligned}}
\def\be{\begin{equation}}
	\def\ee{\end{equation}}
\def\bearr{\begin{eqnarray}}
	\def\eearr{\end{eqnarray}}

\def\l{\left}
\def\r{\right}

\usepackage{tikz}

\usepackage{xcolor}

\hypersetup{
	colorlinks=true,
	linkcolor=blue,
	citecolor=blue,
	urlcolor=blue
}

\newcommand{\showcomments}{\long\def\comm##1\commend{##1}}

\showcomments
\begin{document}
	
\title{Quantum sensing magnonic
number states using a bosonic mode as the probe}

\author{Bashab Dey}
\email{bashab.dey@rptu.de}
\affiliation{Department of Physics and Research Center OPTIMAS, Rheinland-Pf\"alzische Technische Universit\"at Kaiserslautern-Landau, 67663 Kaiserslautern, Germany}

\author{Sonu Verma}
\affiliation{Department of Physics and Research Center OPTIMAS, Rheinland-Pf\"alzische Technische Universit\"at Kaiserslautern-Landau, 67663 Kaiserslautern, Germany}
\affiliation{Institute for High Pressure, Department of Physics, Hanyang University, Seoul 04763, Republic of Korea}

\author{Mathias Weiler}
\affiliation{Department of Physics and Research Center OPTIMAS, Rheinland-Pf\"alzische Technische Universit\"at Kaiserslautern-Landau, 67663 Kaiserslautern, Germany}

\author{Akashdeep Kamra}
\affiliation{Department of Physics and Research Center OPTIMAS, Rheinland-Pf\"alzische Technische Universit\"at Kaiserslautern-Landau, 67663 Kaiserslautern, Germany}

\begin{abstract}    
Sensing  number states of a magnonic mode has been accomplished using a superconducting qubit by realizing an effective dispersive interaction between the two systems. Here, we theoretically demonstrate that a seemingly classical bosonic mode can be utilized as a probe for  resolving the number states of a magnon mode, while outperforming a qubit in various regards as the sensor. Considering another magnon mode in an antiferromagnet as the probe mode, we delineate the required dispersive coupling emerging directly from antiferromagnetic exchange interaction. When a phonon is used as the probe mode, we derive the effective dispersive coupling emerging from the lowest-order nonlinear magnon-phonon interactions. Our  two considered examples provide the general design principles for identifying and utilizing a bosonic probe mode for sensing magnonic superpositions in a physical platform of interest. 
\end{abstract}

\maketitle

%-----------------------------------Intro-----------------------------------------------%

\section{Introduction}  
Magnons are the low-energy quasiparticle excitations of an ordered magnet. Due to their long lifetime and strong coupling with photons, phonons and qubits, magnons can offer a promising platform for quantum computation \cite{Andrianov2014,Hetenyi2022,Terhal2020,Bejarano2024,Yuan2022} and technologies \cite{Laucht2021}. The experimental observation of superpositions of magnon number states \cite{Quirion2017,Xu2023_magnon} and magnon-qubit entanglement \cite{Xu2024_bellstate} has further propelled the research in quantum magnonics \cite{Yuan2022}. While the classical uniform magnon mode can be detected using techniques such as magnetic resonance spectroscopy \cite{Yager1947,Johnson1959}, time-resolved magneto-optical Kerr effect \cite{Beaurepaire1996}, spin pumping and inverse spin Hall effect \cite{Tserkovnyak2002}, resolving the superpositions of its number states has required probing via a qubit \cite{Tabuchi2015, Quirion2017,Quirion2019,Quirion2020,Wolski2020,Rani2025,Dols2024,Xu2023_magnon}. %Achieving a dispersive coupling of the form $\chi \alpha^\dagger \alpha \sigma_z$ with sufficiently large coupling strength $\chi$, where $\alpha$ is the magnon annihilation operator and $\sigma_z$ the qubit operator, is the key to {\color{magenta} sensing these superpositions}.
Spectroscopy of the superconducting qubit enables the superposition sensing through magnon number-dependent shifts in the former's frequency endowed by a strong dispersive interaction \cite{Quirion2017,Faria1999}. Similar interaction  has also been engineered between a photonic mode and a qubit \cite{Schuster2007,Gambetta2006}, a phononic mode and a qubit \cite{Arrangoiz-Arriola2019,Sletten2019,Wollack2022,vonLuepke2022,Lee2023,Cleland2024}, two photonic modes \cite{Milburn_1983,Imoto_1985,Munro_2005,Helmer_2009,Holland2015} or two phononic modes \cite{Roos_2008, Ding_2017} to study their population statistics. 

 Recent breakthroughs in quantum sensing using superconducting \cite{Degen2017,PatelDesai2025,Fink2024,Kakuyanagi2023,Danilin2024} or spinful defect qubits \cite{Casola2018,Xu2023,Simon2022, Sar2015,Page_2019,Dolgirev2022,Chatterjee2022,Bhattacharyya2024,Melendez2025,Machado2023} are enabling unprecedented access to phenomena in solid-state systems. To unleash the full potential of this sensing, two questions take a central stage. First, to what extent is the quantum nature of the qubit fundamental in the sensing protocols? Second, given that good qubits are harder to realize in broad settings, can we achieve similar sensing using some widely available bosonic modes, such as photons or phonons, suited to the specific platform?

\begin{figure}[tbh]
	\centering
	\includegraphics[width=\columnwidth]{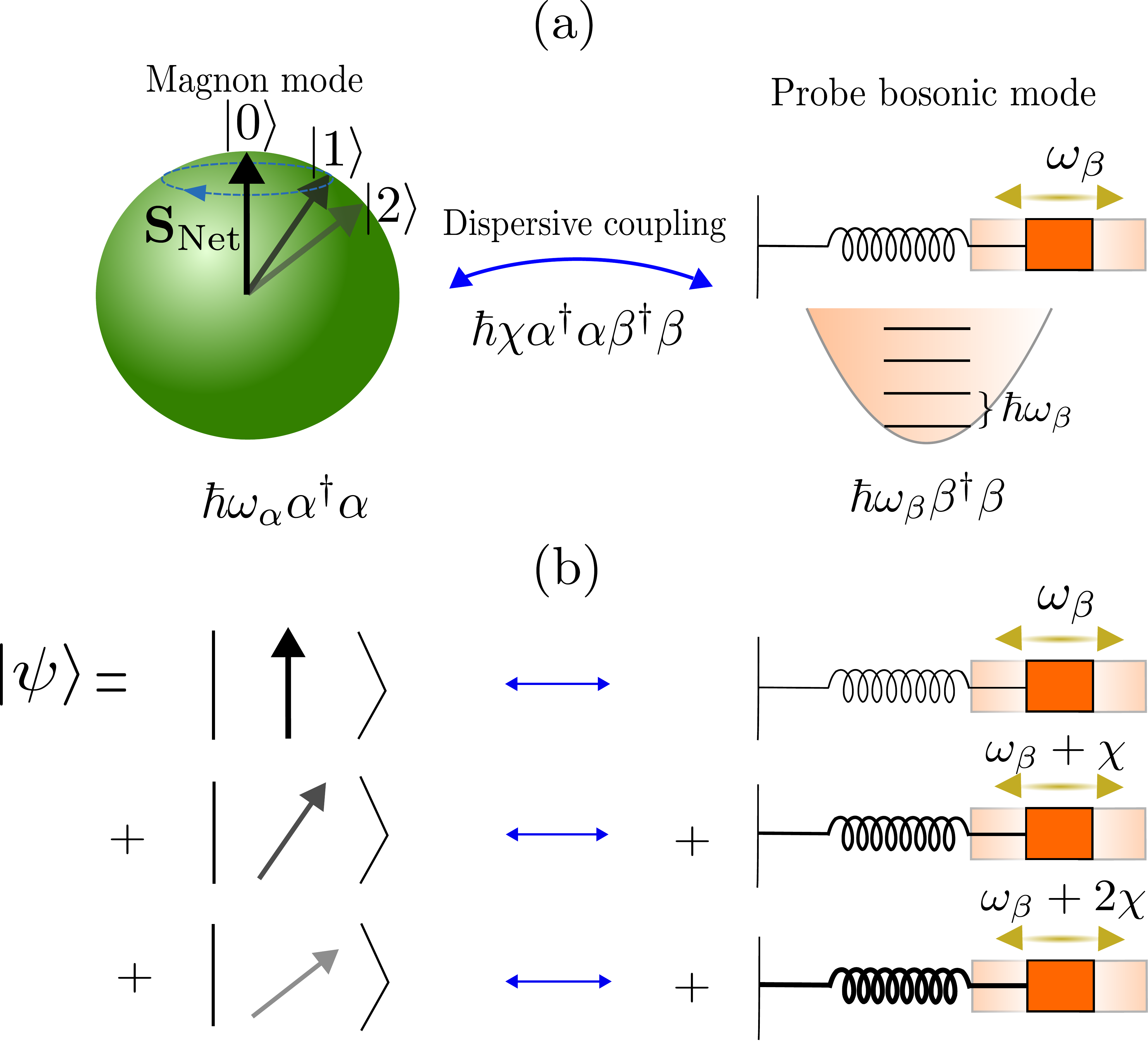}
	\caption{(a) Schematic depiction of the setup. A uniform magnon mode, with annihilation operator $\alpha$, is depicted by an arrow representing the total spin ${\bf S}_\text{Net}$. The tilted arrows represent the excited magnon number states. The probe is a generic harmonic oscillator mode, represented by annihilation operator $\beta$, which interacts with the magnon mode via a dispersive interaction $\hbar \chi \alpha^\dagger \alpha \beta^\dagger \beta$. (b) This dispersive interaction makes the probe mode equivalent spring stiffness and, thus, frequency depend on the magnon number state: $\omega_\beta,\omega
		_\beta+ \chi, \omega_\beta +2\chi...$ for a superposition of the magnon number states $|0\rangle, |1\rangle, |2\rangle,\dots$.}
	\label{f1}
\end{figure}

The community continues to face challenges in sensing quantum magnonic states using a superconducting qubit, with only a few groups having achieved that in the past decade \cite{Xu2023_magnon,Quirion2017,Rani2025}.
Furthermore, the employed architecture requires cryogenic temperatures and a cavity mode to accomplish the desired dispersive magnon-qubit coupling.
In this Letter, we theoretically demonstrate an avenue for employing a wide
range of platforms towards investigating quantum nature of magnons.
%\ak{\sout{,thereby addressing a critical problem being faced by the field}}.}
We theoretically demonstrate that a simple harmonic/bosonic mode can be used as a probe to resolve the magnon number states. This is achieved via a dispersive cross-Kerr coupling $\hbar\chi \alpha^\dagger \alpha \beta^\dagger \beta$ between the magnon (annihilation operator $\alpha$) and probe bosonic (annihilation operator $\beta$) modes which makes the spectroscopically measured probe frequency depend on the magnon number state (Fig.~\ref{f1}). Considering an antiferromagnetic magnon mode as the probe, we show that the required dispersive coupling is {\it direct} and results from the exchange interaction in the antiferromagnet. This is in contrast to the case of a superconducting qubit, where the dispersive coupling was not direct, but achieved as a higher order perturbation theory effect \cite{Quirion2017}. However, to address a broader range of platforms such as a photon or a phonon as the probe mode, we also derive an {\it effective} dispersive coupling from the lowest order three-particle interactions between the two modes. We further find that the unbounded energy spectrum of the probe bosonic mode enables it to outperform the qubit as superposition sensor under non-ideal conditions, such as higher temperatures and stronger coherent drives employed for quasi-classical probe spectroscopy.

%-----------------------------------Dispersive coupling-----------------------------------------------%
\section{Direct dispersive coupling between two magnon modes} 
The exchange interaction has recently been shown to cause a direct dispersive coupling between a spin qubit and the magnon mode to be sensed~\cite{Akash_2023,Akash_2024}. Thus, we may anticipate a similar exchange-induced coupling between the two uniform magnon modes in an antiferromagnet (AFM), with one of them serving as the probe. With this motivation, we first examine an easy-axis anisotropy AFM with $\hat{e}_z$ as the easy axis. In the presence of an external magnetic field  ${\bf H_0}=H_0 \hat{e}_z$, the Hamiltonian of the AFM is given by $\mathcal{H}=\mathcal{H}_\text{E} + \mathcal{H}_\text{A} + \mathcal{H}_\text{Z}$ with the Heisenberg exchange contribution, $\mathcal{H}_\text{E} = \frac{J}{\hbar^2}\sum_{i}^{}\sum_{\boldsymbol{\delta}}^{}{\bf S}_\text{A}({\bf r}_i) \cdot {\bf S}_\text{B}({\bf r}_i+\boldsymbol{\delta}),$ the easy-axis anisotropy $\mathcal{H}_\text{A} = -\frac{K}{\hbar^2}\l(\sum_{i} [S_\text{A}^z({\bf r}_i)]^2 + \sum_j [S_\text{B}^z({\bf r}_j)]^2\r),$
and the Zeeman term $\mathcal{H}_\text{Z}=-\gamma \mu_0 H_0 \l(\sum_i S_\text{A}^z({\bf r}_i) + \sum_j S_\text{B}^z({\bf r}_j)\r).$ Here, $J>0$ is the strength of antiferromagnetic exchange interaction, $K>0$ parameterizes easy-axis anisotropy and $\gamma$ is the gyromagnetic ratio. Indices $i$ and $j$, representing sites on sublattices A and B respectively, span the entire set of $N$ unit cells. The summation $\sum_{\boldsymbol{\delta}}$ in the exchange interaction accounts for all $q$ nearest neighbor sites. The ground state of this system is an antiferromagnetically ordered state with the N\'eel vector oriented along the \( z \)-axis. Employing Holstein-Primakoff transformations \cite{Holstein1940, Hamer1992, Kamra2019}, the Hamiltonian for uniform magnon (${\bf k}=0$) modes is obtained as
\begin{equation}\label{Ham}
    \mathcal{H}_0=\hbar \omega_\alpha \alpha^\dagger \alpha + \hbar \omega_\beta \beta^\dagger \beta +\hbar\chi \alpha^\dagger \alpha \beta^\dagger \beta,
\end{equation}
where $\alpha$ and $\beta$ are the annihilation operators of the eigenmodes [See Supplemental material (SM) \cite{SM} for details]. The bare magnon frequencies are $\omega_{\alpha(\beta)}= \omega_0 + (-)\mu_0 \gamma H_0$, where  $\hbar\omega_0 = 2S\sqrt{K(Jq+K)}$, $S$ is the spin length and $\chi = -Jq/\hbar N$ the dispersive coupling strength. The dispersive coupling is obtained by retaining terms up to the fourth order in the magnon operators and does not require a specific regime of detuning for its validity. Since the dispersive coupling commutes with the uncoupled magnon modes, the energy eigenstates of (\ref{Ham}) are $|n_\alpha\rangle |n_\beta\rangle$ with energies
\begin{equation}
E_{n_{\alpha},n_{\beta}}= \hbar \omega_{\alpha} n_{\alpha} + \hbar( \omega_{\beta} + \chi n_{\alpha}) n_\beta, 
\end{equation}
where $n_\alpha$ and $n_\beta$ are the magnon numbers of the two modes.
%Due to the dispersive interaction, the frequency of one mode depends on the magnon number of the other. 
Considering $\beta$ as the probe bosonic mode, the dispersive coupling manifests itself as the multivalued resonant frequency of the probe oscillator, $\omega_\text{res}^{n_\alpha}=\omega_\beta+ \chi n_\alpha$
when the magnon mode $\alpha$ has a statistical distribution of the magnon number states $|n_\alpha\rangle$ [Fig. \ref{f1}.(b)].\\

%-----------------------------------Spectroscopy-----------------------------------------------%

\section{Resolving magnonic superpositions} %The probe mode frequency can be determined in various ways. For concreteness and theoretical convenience, 
We consider the probe mode to be driven coherently and the probe response is recorded as a function of the drive frequency. When the magnon mode is in a superposition state $|\psi\rangle=\sum_{n_\alpha}^{}c_{n_\alpha} |n_\alpha \rangle$, the  response of the probe mode should peak at a drive frequency of $\omega=\omega^{n_\alpha}_\text{res}$ with $n_\alpha=0,1,2\dots$, the peak heights being proportional to the occupation probabilities $P_{n_\alpha}\equiv|c_{n_\alpha}|^2$ of the corresponding magnon number states. This is analogous to qubit spectroscopy used for the sensing of photon-, magnon- and phonon-number statistics \cite{Schuster2007,Quirion2017,Arrangoiz-Arriola2019,Akash_2023}.

\begin{figure}[t]
	\centering
	\vspace{1cm}
	\includegraphics[width=8.5cm]{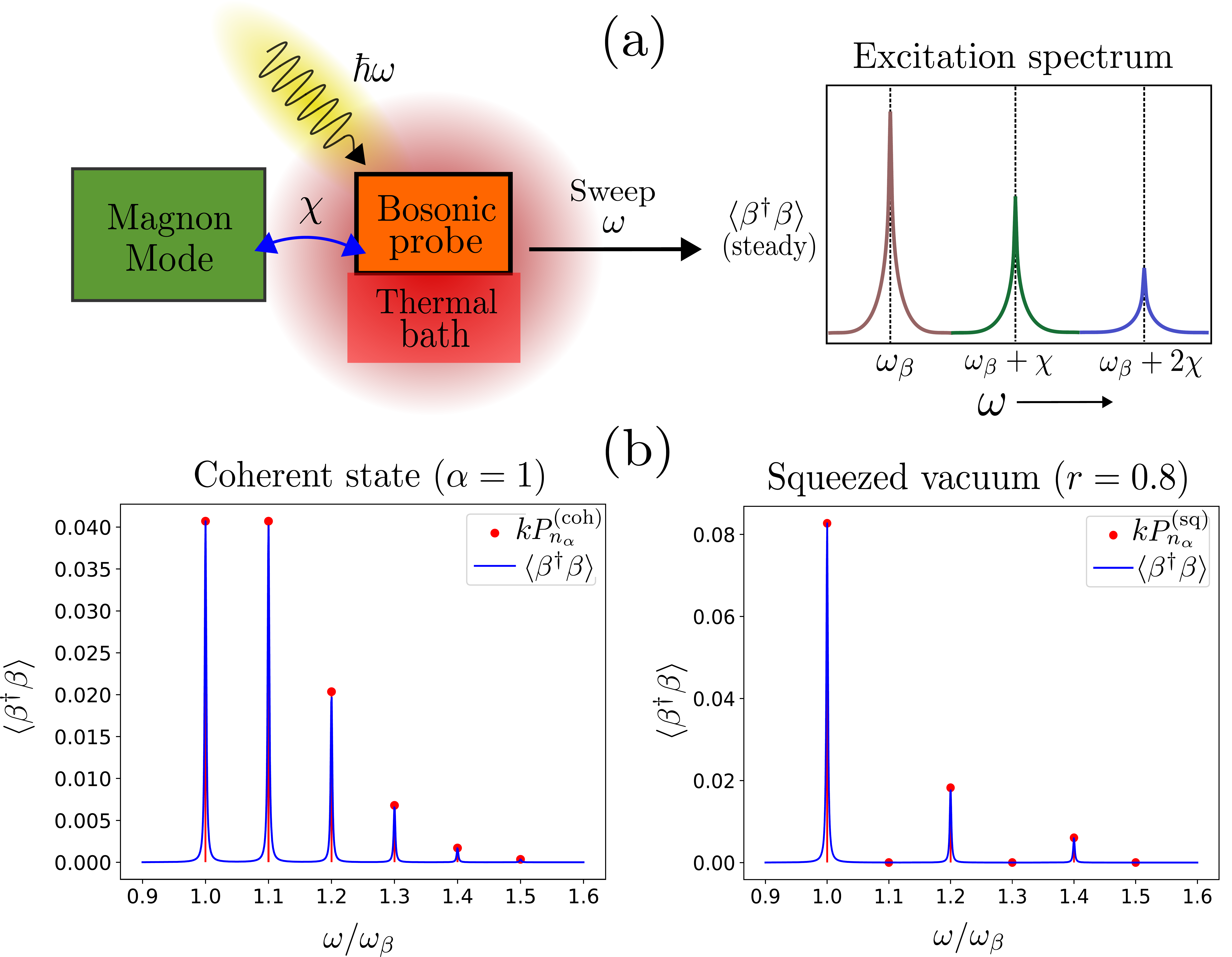}
	\caption{(a) Schematic depiction of the probe mode spectroscopy. The bosonic probe is coupled to a thermal bath at temperature $T$ and coherently driven. The drive frequency $\omega$ is swept and the steady state probe mode excitation $\langle\beta^\dagger \beta\rangle$ is plotted as function of $\omega$. Peaks appear at multiple frequencies corresponding to unique number states in the magnonic  distribution being sensed. (b) Simulation data of the spectroscopy when the magnon mode is in a coherent state (left) with $\alpha=1$ and squeezed vacuum (right). With $\alpha$ as the coherent state amplitude and $r$ the squeezing factor, the parameters used in the simulation are: $\alpha=1.0,~ r=0.8, ~\chi/\omega_\beta=0.1,~ d/\hbar\omega_\beta=0.001, \kappa/\omega_\beta=0.003$ and  $k_BT/\hbar\omega_\beta =0.01$. The blue curves showing the numerically simulated spectroscopic peaks are in good agreement with the exact probability distributions of the magnon number states shown by the red dots.}
	\label{f2}
\end{figure}

We demonstrate the sensing protocol by numerically simulating the corresponding spectroscopy \cite{Akash_2023,Akash_2024}. The setup consists of the magnon mode dispersively interacting with the probe bosonic mode, which in turn is coupled to a thermal bath at temperature $T$ [Fig. \ref{f2}.(a)]. %A thermal bath is chosen to check the efficiency of the probe at finite temperatures. 
A coherent drive $\mathcal{V}(t)= d\cos\omega t (\beta + \beta^\dagger)$ of strength $d$ is applied to the probe.  The time evolution of the system is studied by numerically solving the Lindblad master equation \cite{Breuer2002, Lindblad1976}
 \begin{equation}\label{Lindblad}
      \frac{d\rho}{dt}=-\frac{i}{\hbar}[\mathcal{H}(t),\rho(t)] + \sum_{n=1,2}^{} C_n \rho(t) C_n^\dagger -\frac{1}{2}\l\{\rho(t), C_n^\dagger C_n\r\}
 \end{equation}
where $\rho(t)$ is the density matrix and $\mathcal{H}(t)\equiv\mathcal{H}_0 + \mathcal{V}(t)$ is the Hamiltonian of the magnon-probe system. The collapse operators $C_1=\sqrt{\kappa(n^\beta_\text{th}+1)} ~\beta$ and $C_2=\sqrt{\kappa~n^\beta_\text{th}}~ \beta ^\dagger$ characterize energy emission to, and absorption from, the bath, respectively. The decay rate of the probe mode is $\kappa$, while its equilibrium thermal occupation is $n^\beta_\text{th}=[e^{\hbar \omega_\beta/k_B T}-1]^{-1}$. The decay in the magnon mode $\alpha$ is implicit in our model as it determines the magnonic distribution $\{P_{n_\alpha}\}$ to be sensed, whose information is contained in $\rho(0)$ \cite{Gambetta2006,Schuster2007,Akash_2023}. The magnon decay considered here is equivalent to the Gilbert damping used in the Landau-Lifshitz framework \cite{Yuan2022_Lindblad}. 
When a steady state is reached upon driving, the probe excitation $\langle \beta^\dagger \beta \rangle \equiv \text{Tr}[\rho  \beta^\dagger \beta]$ peaks at the resonant frequencies $\omega=\l\{\omega_\text{res}^{n_\alpha}\r\}$ [Fig. \ref{f2}(a)]. The separation between the peaks is $|\chi|$. At $T=0$, the excitation at $\omega=\omega_\text{res}^{n_\alpha}$ can be written as $\langle\beta^\dagger \beta \rangle_{n_\alpha} = k P_{n_\alpha}$, where $k$ is a proportionality constant depending on $d$ and $\kappa$. %A more rigorous inclusion of dissipation in the mode to be sensed is beyond the scope of the present work and will be addressed in future efforts aimed at system-specific details. 

As a demonstrative example, we present the simulation data for the magnon mode being in two states -- coherent state (classical) and squeezed vacuum (nonclassical), with the bath at zero temperature [Fig.~\ref{f2}(b)]. Here, $\{P^\text{(coh)}_{n_\alpha}\}$ and $\{P^\text{(sq)}_{n_\alpha}\}$ are the exact probability distributions of the number states in the coherent state and squeezed vacuum, respectively \cite{Gerry2005}. Upon extracting the proportionality constant $k$ from the normalization condition (see SM \cite{SM}), we find that the simulated peaks $\langle \beta^\dagger \beta\rangle_{n_\alpha}$ are in good agreement with $k P^\text{(coh)}_{n_\alpha}$ and $k P^\text{(sq)}_{n_\alpha}$ shown via red dots in the figure. This validates the anticipated resolution of magnon number state superpositions using a bosonic mode as the probe.  The simulations have been performed using the QUTIP package \cite{Johansson2012,Johansson2013}. %The simulation results are independent of $\omega_\alpha$ and $\omega_\beta$ values. 

%-----------------------------------Temperature effects-----------------------------------------------%

\begin{figure*}
		\centering
\includegraphics[width=0.75\textwidth]{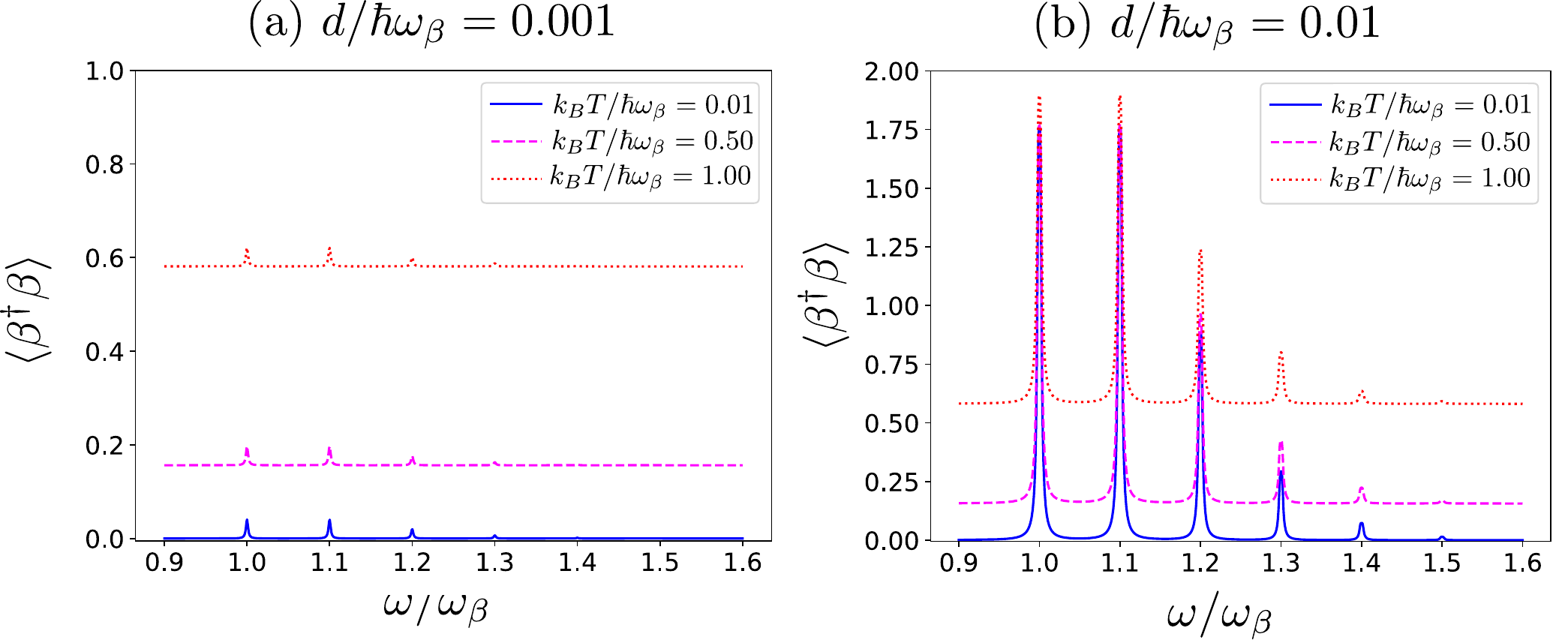}
		\caption{ Simulation results of the probe spectroscopy at different temperatures for a (a) weak drive ($d/\hbar\omega_\beta=0.001$) and (b) a strong drive ($d/\hbar\omega_\beta=0.01$). With increase in temperature, the mean thermal occupation $n_\text{th}^\beta$ of the probe mode rises. For weak driving, the spectroscopy peaks are small as compared to $n_\text{th}^\beta$, leading to poor experimental detectability. A stronger drive enhances the peaks making them detectable.}
		\label{f3}
	\end{figure*}

\section{Resolvability of the magnon number states at finite temperatures}
In contrast with the two-level qubit, a bosonic mode at finite temperature occupies several number states. Thus, one may expect finite temperature to be more detrimental to the bosonic mode as a sensor. We examine this question now and find instead that the bosonic mode driven with a coherent drive is more robust than the qubit for sensing the desired magnonic superpositions.

In Fig.~\ref{f3}, we show the spectroscopy simulation data at different temperatures for weak and strong driving.  Away from resonance, the probe mode has the mean thermal excitation $n_\text{th}^\beta$, which determines the baselines in spectroscopy plots. At a resonance, the drive additionally excites the probe mode creating a Glauber state \cite{Glauber1963} which represents the classical mode of oscillation. So, the magnitude of this extra excitation should be independent of temperature and proportional to the driving power \cite{Drummond_1980, Gardiner2004}. This is indeed reflected in the spectroscopy as peak heights, measured from the baseline, do not change with temperature [Fig. \ref{f3}(a)] and still measure the desired magnon number statistics. The consequent net occupation of the probe at a resonance $\langle \beta^\dagger \beta\rangle_{n_\alpha}$ is derived as (see SM \cite{SM} for details):
\begin{equation}\langle \beta^\dagger \beta\rangle_{n_\alpha}= k P_{n_\alpha} + n_\text{th}^\beta,\end{equation}
%\begin{equation}\langle\beta^\dagger\beta\rangle_\text{ss}=\sum_{n_\alpha}P_{n_\alpha}\l(\frac{d^2}{\gamma^2+4[(\omega_\beta+\chi n_\alpha)-\omega]^2}\r)+n_\text{th}\end{equation}
where $k=d^2/\hbar^2\kappa^2$ and $k P_{n_\alpha}$ is the peak height from the baseline. At high temperatures and weak driving, the peak heights are much smaller than $n_\text{th}^\beta$ which can limit their detectability. This issue is circumvented by driving the probe stronger so that the peaks are high enough [Fig. \ref{f3}(b)] without compromising their  proportionality to the magnon number statistics being sensed.

\begin{figure}[t]
		\centering
        \vspace{0.5cm}
\includegraphics[width=8.5cm]{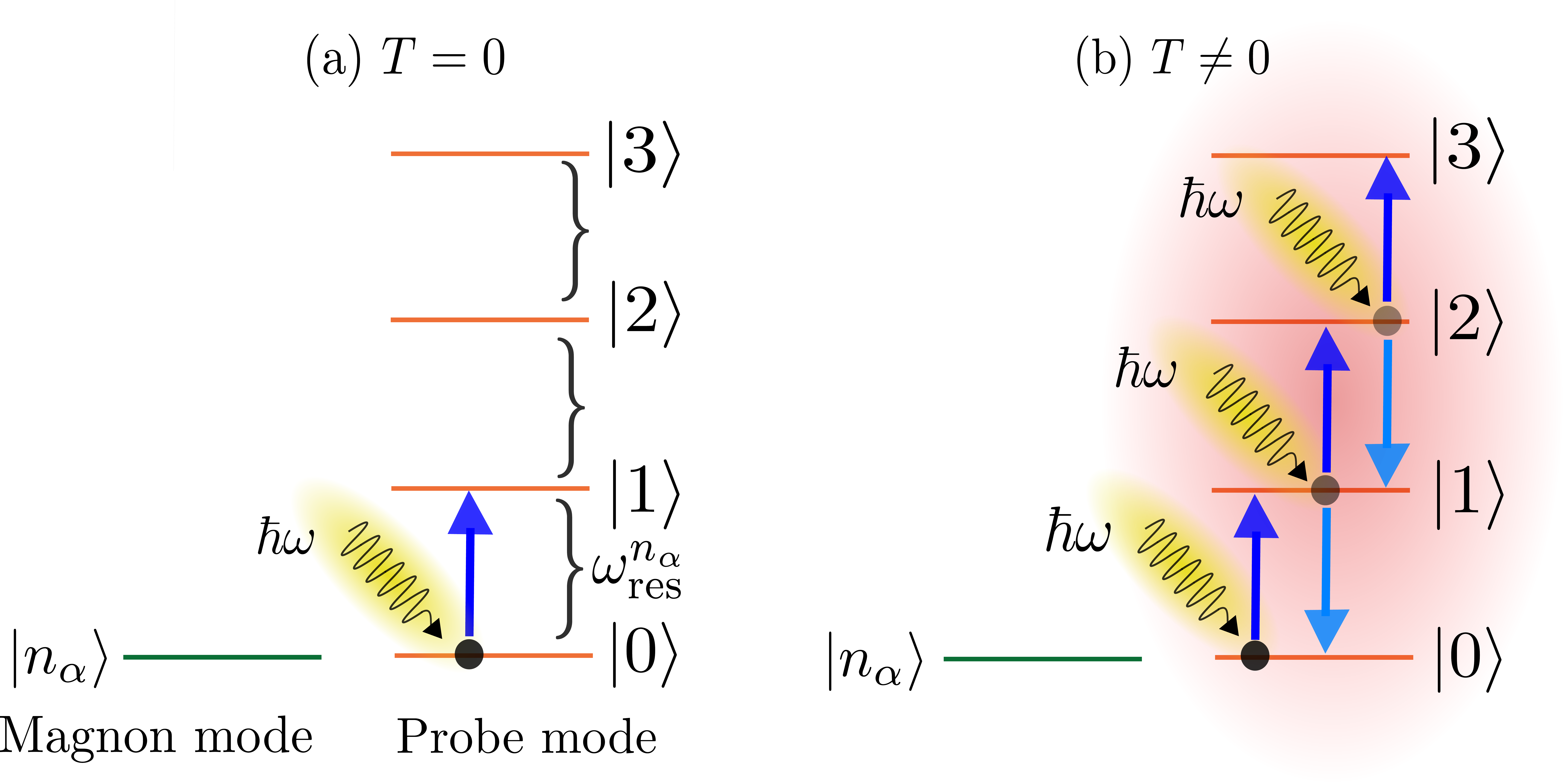}
		\caption{Schematic depiction of the quantum transitions in the probe mode. (a) At $T=0$, a photon is resonantly absorbed at $\omega=\omega_\text{res}$ causing a transition $|n_\alpha\rangle|0\rangle \to |n_\alpha\rangle|1\rangle $. (b)  At $T\neq0$, the probe mode has a thermal distribution of the boson number states. Stimulated absorptions (up-arrow) and emissions (down-arrow) occur between adjacent levels. The probe still remains a simple harmonic oscillator as the energy levels remain equally spaced i.e. $\Delta=\hbar\omega_\text{res}^{n_\alpha}$ despite the superposition of the magnon number states. This makes the bosonic probe a robust sensor of the magnon number statistics even at finite temperatures.}
		\label{f4}
	\end{figure}

The resonant peaks and their immunity to temperature can be understood intuitively in terms of the quantum transitions that occur between the energy levels $|n_\alpha\rangle |n_\beta\rangle$ of the system. The energy difference between any two levels is
 \begin{equation}
 \begin{aligned}
     &E_{n_\alpha^\prime, n_\beta^\prime} - E_{n_\alpha, n_\beta}\\ &= \hbar \big[\omega_\alpha (n_\alpha^\prime-n_\alpha) + \omega_\beta (n_\beta^\prime-n_\beta)+ \chi (n_\alpha^\prime n_\beta^\prime -n_\alpha n_\beta)\big]
 \end{aligned}
 \end{equation}
The drive $ \sim  f(\omega t)(\beta + \beta^\dagger)$ only couples levels for which $n_\alpha^\prime=n_\alpha$ and $n_\beta^\prime=n_\beta\pm1$. The corresponding energy difference is
\begin{equation}\label{freq}
    E_{n_\alpha, n_\beta\pm1} - E_{n_\alpha, n_\beta}=\pm \hbar(\omega_\beta+\chi n_\alpha)=\pm \hbar \omega_\text{res}^{n_\alpha},
\end{equation} 
which is independent of the probe mode occupation $n_{\beta}$. Hence, within linear response, the drive causes resonant transitions $|n_\alpha\rangle |n_\beta\rangle \to |n_\alpha\rangle |n_\beta\pm1\rangle$ when $\omega=\omega_\text{res}^{n_\alpha}$. %The $+~(-)$ sign corresponds to stimulated absorption (emission) of a photon. Equation (\ref{freq}) tells that the states $|n_\alpha\rangle |n_\beta\rangle$ with a fixed $n_\alpha$ and different $n_\beta ~(=0,1,2,\dots)$ have equal level-spacing $\Delta=\hbar\omega_\text{res}^{n_\alpha}$. Thus, for a given resonance, the probe is still a linear harmonic oscillator despite the superposition of the magnon number states. 
A schematic of the quantum transitions at a resonance is shown in Fig.~\ref{f4} for zero and finite temperatures of the probe.  At $T=0$, the probe mode is in its ground state and a resonant transition $|n_\alpha\rangle|0\rangle \to |n_\alpha\rangle|1\rangle $ occurs when $\omega=\omega_\text{res}^{n_\alpha}$ [Fig. \ref{f4}(a)]. At $T\neq0$, the probe levels are thermally populated. So, both stimulated absorptions $|n_\alpha\rangle |n_\beta\rangle \to |n_\alpha\rangle|n_\beta+1\rangle$ and emissions  $|n_\alpha\rangle |n_\beta\rangle \to |n_\alpha\rangle|n_\beta-1\rangle$ occur upon driving at the same frequency $\omega_\text{res}^{n_\alpha}$, as shown by up- and down-arrows, respectively, in Fig.~\ref{f4}(b).  %Due to equal level-spacings, the resonant frequencies of stimulated absorption and emission are identical i.e. $\omega_\text{res}^{n_\alpha}$. 
Since the resonant frequency is independent of the boson number $n_\beta$, it is not modified by the thermal distribution of the probe number states and the spectroscopy protocol for sensing the magnon number statistics in the $\alpha$ mode remains valid.

Due to its two-level nature, a qubit is constrained with respect to its maximum excitation \cite{Breuer}. The latter needs to remain small for the spectroscopy to reveal the desired   magnonic statistics  \cite{Akash_2023}. This constraint gets lifted when using a bosonic mode as the probe since it has an infinitely large Hilbert space thereby enabling effective sensing at higher temperatures. A detailed simulation and comparison presented in the SM~\cite{SM} corroborates this superiority of the bosonic mode over the qubit as a sensor. 

%-----------------------------------Effective coupling-----------------------------------------------%

\section{Effective dispersive interaction between a magnon and phonon mode} Unlike the case of an AFM, there may not be a direct dispersive interaction between the magnon mode to be sensed and a generic probe bosonic mode. However, an effective dispersive interaction could be engineered in a wide range of systems that do not host direct dispersive coupling. For example, the linear  coherent coupling between a qubit and bosonic mode results in an effective dispersive coupling in the detuned limit~\cite{Savage1990, Faria1999, Blais2004,Zueco2009}.%, which enables dispersive readout of a qubit state \cite{Wallraff2004, Liu2005, Lupascu2006} and quantum nondemolition measurement of magnon- \cite{Quirion2017} and photon-number statistics \cite{Schuster2007}.

Magnomechanical interactions in a ferromagnet also result in magnon-phonon couplings \cite{Abrahams1952} that are linear $\sim\alpha^\dagger \beta$ + H.c. \cite{Potts2021} or quadratic $\propto \alpha^{\dagger2} \beta, \alpha^2 \beta, \alpha^\dagger\alpha \beta $ + H.c. \cite{Rueckriegel2014, Ballestero2020_prb, Bittencourt2023,kuenstle2025} in the magnon operators, where $\alpha$ and $\beta $ are the annihilation operators of the magnon and phonon modes respectively. Among the quadratic ones, only the parametric coupling  $\sim  \alpha^\dagger\alpha \beta$ + H.c. is considered relevant in the far-detuned limit \cite{Zhang2016,Li2018,Li2019,Ballestero2020_prl,Ballestero2020_prb,Bruehlmann2026}. In our analysis, %aimed at achieving the desired dispersive coupling between the two bosonic modes, 
we find that the linear $\sim  \alpha^\dagger \beta$ + H.c. and parametric couplings $\sim  \alpha^\dagger \alpha \beta$ + H.c. do not generate any dispersive interaction in the far-detuned limit (see SM \cite{SM}) while the parametric-conversion $\sim\alpha^{\dagger2} \beta $ + H.c. and the counter-rotating terms $\sim \alpha^{2} \beta $ + H.c. do. %indeed give rise to the desired effective dispersive interaction. 
To establish design principles, we consider the most general three-particle interaction:
\begin{equation}\label{cubic}
\begin{aligned}
     V_\text{int}&= \hbar \chi_1 (\alpha^\dagger \beta^2+\alpha \beta^{\dagger2}) + \hbar \chi_2 (\alpha \beta^2+\alpha^\dagger \beta^{\dagger2})\\
     &+ \hbar \chi_3 (\alpha^2 \beta^\dagger+\alpha^{\dagger2} \beta) + \hbar \chi_4 (\alpha^2 \beta +\alpha^{\dagger2} \beta^\dagger).
\end{aligned}
\end{equation}
In the far-detuned limit, a Schrieffer-Wolff transformation \cite{Schrieffer1966,Landi2024} on the magnon-phonon Hamiltonian $H_\text{mp}=\hbar \omega_\alpha \alpha^\dagger \alpha + \hbar \omega_\beta \beta^\dagger \beta + V_\text{int}$ gives an effective Hamiltonian 
\begin{equation}\label{cubiceff}
    \l[H_\text{mp}\r]_\text{eff}=\hbar \omega_\alpha \alpha^\dagger \alpha + \hbar \omega_\beta \beta^\dagger \beta + \hbar \chi_\text{eff}\alpha^\dagger\alpha\beta^\dagger \beta,
\end{equation}
where
\begin{equation}\label{chieff}
    \chi_\text{eff}=-\frac{4 \chi_1^2}{2\omega_\beta-\omega_\alpha}-\frac{4 \chi_2^2}{2\omega_\beta+\omega_\alpha}+\frac{4  \chi_3^2}{\omega_\beta-2\omega_\alpha}-\frac{4 \chi_4^2}{\omega_\beta+2\omega_\alpha}
\end{equation}
is the effective dispersive strength (see SM \cite{SM} for details). While we mention a phonon mode as a probe here, our analysis is generic and applies to any bosonic mode that bears the required nonlinear coupling with the magnon mode to be sensed.

%-----------------------------------Physical realizations-----------------------------------------------%

\section{Physical realization} Considering the AFM with exchange-mediated direct dispersive coupling as the platform, hematite in its easy-plane anisotropy phase hosts a high-frequency magnon mode in sub-THz regime and a low-frequency one in few GHz range \cite{Morrish1995,Boventer2021,WangH2021,Chen2025,Hamdo2025}. One may utilize the latter for sensing the magnonic superpositions in the former at moderately high temperatures. %A wide temperature range is expected to appear as nearly zero to the high-frequency mode. 
Since we show that a finite temperature is not a major problem for the probe mode, the desired sensing of the superpositions is feasible.   The design principles for using an AFM magnon or phonon mode as the probe have been detailed in the SM \cite{SM}. We deduce that the direct dispersive strength $\chi\propto1/V$, where $V$ is the volume of the sample. Since the linewidth of the low-frequency resonance in hematite can be several tens of MHz \cite{Boventer2021}, the spectroscopic peaks should be resolvable for nanometer-sized samples \cite{Zysler2004,Vamvakidis2013,Yang2015,Foner1963} where the dispersive shift $\chi/2\pi$ can be in the low or sub-GHz regime. %The design guidance for hematite has been detailed in the SM \cite{SM}. \sout{For larger AFMs, $\chi$ may be enhanced by making thin films} \cite{Lui1986,Cortie2012,Yuan2016,Appel2019} \sout{which would reduce $N$ along the thickness of the film}.
Considering a phonon mode as the probe, we show that the strength of the parametric down-conversion type coupling $\chi_3$ in Eq.(\ref{cubic}) varies as $1/ \sqrt{V}$. We estimate that $\chi_3$ can go up to few tens of MHz for thin ferromagnetic films \cite{kuenstle2025,Hwang_Otani_2025}. Choosing a moderately large detuning of $\sim 6\chi_3$, the dispersive shift $\chi_\text{eff}/2\pi$ can be of the order of MHz.  Using a phononic mode as the probe for ferromagnetic magnons offers more control over the effective dispersive coupling as the detuning can be optimally chosen for maximizing $\chi_\text{eff}$.

Apart from the dispersive term, Kerr-like nonlinearities $(\sim \alpha^{\dagger^2} \alpha^2, \beta^{\dagger2} \beta^2$), frequency-pulling $(\sim \alpha^\dagger \alpha, \beta^\dagger \beta)$ and mixed terms  $(\sim \alpha^{\dagger^2}\beta^\dagger \beta,$ etc.) also appear in the full Hamiltonians of Eqs.~(\ref{Ham}) and (\ref{cubiceff}). The Kerr nonlinearity \cite{Drummond_1980,Candeloro2021,Asjad2023} renders the driven probe nonclassical due to which the resolution of the superpositions is affected at finite temperature, as detailed in the SM \cite{SM}.  The Kerr nonlinearity arises in the probe mode ($\beta$) due to the nonlinear couplings $\propto \chi_1$ and $\chi_2$ in Eq. (\ref{cubic}). %A suitable system design that suppresses $\chi_1$ and $\chi_2$  can reduce the Kerr effects.
Choosing the operating point in the frequencies that maximizes the contribution of $\chi_3$ or $\chi_4$ in $\chi_\text{eff}$ [Eq. (\ref{chieff})] automatically suppresses the corresponding contributions from $\chi_1$ or $\chi_2$ (see SM \cite{SM}). The frequency-pulling term creates only a global shift in the frequencies of the two modes, which does not affect the simulation results.

To conclude, we have demonstrated how the probability distribution of the number states of a magnon eigenmode can be sensed using a bosonic probe by coupling it to the magnon mode dispersively. The spectroscopy works well even at high temperatures and stronger driving of the probe mode, which offers a significant advantage over the use of a qubit for this sensing. The dispersive coupling, necessary for the sensing, can be direct or effective. For selected platforms, direct dispersive coupling can be dominant,
while for some others, effective dispersive coupling may be more suitable. We discussed two concrete examples of the probe mode -- a magnon mode of an AFM and a phonon mode of a ferromagnet, but any bosonic mode can be used. Our analysis lays the groundwork for identifying and utilizing the optimal bosonic probe modes in a given system of interest for quantum manipulation and sensing of superpositions. 

\textit{Note added.---} During the review process, a related work investigating and demonstrating the cross-Kerr interaction with magnons has appeared~\cite{Fruy2026}.

\begin{acknowledgments}
	We thank Takis Kontos for informing us about his related unpublished work and the subsequent discussions. We thank the German Research Foundation (DFG) for funding via Spin+X TRR 173-268565370, projects A13 and B13. S.V. acknowledges support from the Brain Pool Program funded by the Ministry of Science and ICT through the National Research Foundation of Korea (RS-2025-25446099).
\end{acknowledgments}

\nocite{Skogvoll2021,Pincus1960}
\bibliography{bibliography} 

\clearpage

\onecolumngrid

\begin{center}
    {\large\bfseries Supplemental Material
    accompanying the manuscript}\\[1ex] {\large\bfseries``Quantum sensing magnon number states using a bosonic mode as the probe''
    }\\[2ex]

    Bashab Dey${}^1$, 
    Sonu Verma${}^{1,2}$,
    Mathias Weiler${}^1$,
    Akashdeep Kamra${}^1$

    \vspace{1ex}

    \small \it
     ${}^1$Department of Physics and Research Center OPTIMAS,
    Rheinland-Pf\"alzische Technische Universit\"at Kaiserslautern-Landau,
    67663 Kaiserslautern, Germany

    \vspace{0.5ex}

    ${}^2$Institute for High Pressure, Department of Physics,
    Hanyang University, Seoul 04763, Republic of Korea
\end{center}

\bigskip

\section*{Contents}
\begin{itemize}
    \item[] \hyperref[sec: derivation]{I. Derivation of direct dispersive interaction between the magnon eigenmodes of an antiferromagnet}
    \item[] \hyperref[sec: simulation]{II. Simulation of the spectroscopy}
    \begin{itemize}
        \item [] \hyperref[subsec: formalism]{A. Formalism}
        \item [] \hyperref[subsec: simulation]{B. Simulation}
        \item [] \hyperref[subsec: resolving]{C. Resolving the magnon number states}
    \end{itemize}
    \item[] \hyperref[sec: comparison]{III. Comparison between qubit and bosonic mode as sensors of magnon number state }
    \item[] \hyperref[effective-dispersive]{IV. Effective dispersive interaction between a magnon and phonon mode }
    \item[] \hyperref[Kerr]{V. Effect of a Kerr nonlinearity in the probe bosonic mode }
    \begin{itemize}
        \item[] \hyperref[subsec: Kerr-simulation]{A. Simulation results }
        \begin{itemize}
            \item[] \hyperref[subsubsec: Kerr-simulation-zeroT-strong]{1.  Zero temperature and strong driving  }
            \item[] \hyperref[subsubsec: Kerr-simulation-highT-weak]{2. High temperature and weak driving   }
            \item[] \hyperref[subsubsec: Kerr-simulation-highT-strong]{3. High temperature and strong driving  }
        \end{itemize}
    \end{itemize} 
    \item[] \hyperref[sec: proportionality]{VI. Proportionality of the spectroscopic peaks to magnonic probability distribution }
    \begin{itemize}
        \item[] \hyperref[subsec: time-dependent]{A. Time-dependent perturbation theory approach }
        \item[] \hyperref[lindblad-section]{B. Lindblad master equation approach  }
    \end{itemize}
    \item[] \hyperref[sec: Design]{VII. Design Principles }
    \begin{itemize}
        \item[]  \hyperref[subsec: afm mode]{A. Antiferromagnetic magnon mode as probe }
   \item[]  \hyperref[subsec: phonon mode]{B. Phonon mode as probe } 
    \end{itemize}
\end{itemize}

\clearpage

\section{Derivation of direct dispersive interaction between the magnon eigenmodes of an antiferromagnet}\label{sec: derivation}

We consider an antiferromagnet (AFM) with $\hat{e}_z$ as the easy axis. On application of a magnetic field ${\bf H_0}=H_0 \hat{e}_z$, the spin Hamiltonian of the AFM is written as \cite{Kamra2019}
\begin{equation}
    \mathcal{H}=\mathcal{H}_\text{E} + \mathcal{H}_\text{A} + \mathcal{H}_\text{Z}
\end{equation}
where
\begin{equation}\label{exchange-ham}
    \mathcal{H}_\text{E}=\frac{J}{\hbar^2}\sum_{i}^{}\sum_{\boldsymbol{\delta}}^{}{\bf S}_\text{A}({\bf r}_i)\cdot {\bf S}_\text{B}({\bf r}_i+\boldsymbol{\delta}),
\end{equation}
\begin{equation}
    \mathcal{H}_\text{A}=-\frac{K}{\hbar^2}\l(\sum_{i} [S_\text{A}^z({\bf r}_i)]^2 + \sum_{j}[S_\text{B}^z({\bf r}_j)]^2\r),
\end{equation}
and
\begin{equation}
    \mathcal{H}_\text{Z}=-\gamma \mu_0 H_0 \l(\sum_{i} S_\text{A}^z({\bf r}_i) + \sum_{j} S_\text{B}^z({\bf r}_j)\r)
\end{equation}

are Heisenberg exchange interaction, easy-axis anisotropy and Zeeman interaction respectively. Here, the constant $J>0$ is the strength of antiferromagnetic exchange interaction, $K>0$ parameterizes easy-axis anisotropy and $\gamma$ is the gyromagnetic ratio. Indices $i$ and $j$, representing sites on sublattices A and B respectively, span the entire set of $N$ magnetic unit cells. The summation $\sum_{\boldsymbol{\delta}}$ in the exchange interaction accounts for all $q$ nearest neighbor sites. The ground state of this system is an antiferromagnetically ordered state with the N\'eel vector oriented along the \( z \)-axis. The exchange interaction can be simplified as

\begin{equation}
\begin{aligned}
    \mathcal{H}_\text{E}=&\frac{J}{\hbar^2}\sum_{i}\sum_{\boldsymbol{\delta}}\l\{\frac{1}{2}\l[S_\text{A}^+({\bf r}_i)S_\text{B}^-({\bf r}_i+\boldsymbol{\delta}) + S_\text{A}^-({\bf r}_i)S_\text{B}^+({\bf r}_i+\boldsymbol{\delta})\r]+S_\text{A}^z({\bf r}_i)S_\text{B}^z({\bf r}_i+\boldsymbol{\delta})\r\}.
\end{aligned}
\end{equation}
where $S^{+(-)}=S_x + (-) i S_y$ is the spin-raising (lowering) operator.
Let $S$ be the spin-length of each spin on the lattice. The spin operators are bosonized using the following Holstein-Primakoff transformations \cite{Holstein1940, Hamer1992, Kamra2019}:
\begin{equation}
    S_\text{A}^+({\bf r}_i)\approx\hbar \sqrt{2S} \l(1-\frac{a_i^\dagger a_i}{4S}\r) a_i,~~~~~~
    S_\text{A}^-({\bf r}_i)\approx\hbar \sqrt{2S}~ a_i^\dagger\l(1-\frac{a_i^\dagger a_i}{4S}\r),~~~~~~
    S_\text{A}^z({\bf r}_i)=\hbar (S-a^\dagger_i a_i).
\end{equation}
\begin{equation}
    S_\text{B}^+({\bf r}_j)\approx\hbar \sqrt{2S}~ b_j^\dagger \l(1-\frac{b_j^\dagger b_j}{4S}\r),~~~~~
    S_\text{B}^-({\bf r}_j)\approx\hbar \sqrt{2S}~ \l(1-\frac{b_j^\dagger b_j}{4S}\r) b_j,~~~~~
    S_\text{B}^z({\bf r}_j)=\hbar (-S+b^\dagger_j b_j),
\end{equation}
Retaining terms upto products of four-magnon operators, we get
\begin{equation}
    \mathcal{H}_\text{E}=-NqJS^2+\sum_{\langle i,j\rangle} \l[J S \l(a_i^\dagger a_i + b_j^\dagger b_j + a_i b_j +a_i^\dagger b_j^\dagger\r) -\frac{J}{4} \l(a_i b_j^\dagger b_j b_j + a_i^\dagger a_i a_i b_j + a_i^\dagger b_j^\dagger b_j^\dagger b_j + a_i^\dagger a_i^\dagger a_i b_j^\dagger\r)-J a_i^\dagger a_i  b_j^\dagger b_j\r], 
\end{equation}
\begin{equation}
    \mathcal{H}_\text{A}=-2KNS^2 + 2KS\l(\sum_i a_i^\dagger a_i  + \sum_j b_j^\dagger b_j \r)-K\l[ \sum_i (a_i^\dagger a_i)^2  + \sum_j (b_j^\dagger b_j)^2 \r],
\end{equation}
and
\begin{equation}
    \mathcal{H}_\text{Z}=\gamma \hbar \mu_0 \mathcal{H}_0\l( \sum_ia_i^\dagger a_i - \sum_i b_j^\dagger b_j \r),
\end{equation}
where $\langle i,j\rangle$ in the first summation implies nearest neighbor interactions. The terms of the Hamiltonian can be regrouped as
\begin{equation}\label{Hreal}
\begin{aligned}
&
    \mathcal{H}=-NS^2(Jq+2K)+ \l[(Jq+2K)S+\gamma \mu_0 \hbar H_0 \r] \sum_i a_i^\dagger a_i +\l[(Jq+2K)S-\gamma \mu_0 \hbar H_0 \r] \sum_j b_j^\dagger b_j\\
    &+JS \sum_{\langle i,j\rangle} ( a_i b_j +a_i^\dagger b_j^\dagger) - J\l[ \sum_{\langle i,j\rangle} \frac{1}{4}\l(a_i b_j^\dagger b_j b_j + a_i^\dagger a_i a_i b_j+ a_i^\dagger b_j^\dagger b_j^\dagger b_j + a_i^\dagger a_i^\dagger a_i b_j^\dagger \r)+ a_i^\dagger a_i  b_j^\dagger b_j\r]-K\l[ \sum_i (a_i^\dagger a_i)^2  + \sum_j (b_j^\dagger b_j)^2 \r].
\end{aligned}
\end{equation}
 We perform the Fourier transformations
\begin{equation}
a_i=\frac{1}{\sqrt{N}}\sum_{\bf k}a_{\bf k} \e^{i{\bf k}\cdot{\bf r}_i}, ~~a_i^\dagger=\frac{1}{\sqrt{N}}\sum_{\bf k}a_{\bf k}^\dagger \e^{-i{\bf k}\cdot{\bf r}_i}, ~~ b_j=\frac{1}{\sqrt{N}}\sum_{\bf k}b_{\bf k} \e^{-i{\bf k}\cdot{\bf r}_j}, ~~b_j^\dagger=\frac{1}{\sqrt{N}}\sum_{\bf k}b_{\bf k}^\dagger \e^{i{\bf k}\cdot{\bf r}_j}.
\end{equation}
to get
\begin{equation}
    \sum_i a_i^\dagger a_i = \sum_{\bf k}a_{\bf k}^\dagger a_{\bf k},
\end{equation}
\begin{equation}
    \sum_i b_j^\dagger b_j = \sum_{\bf k}b_{\bf k}^\dagger b_{\bf k},
\end{equation}
\begin{equation}
    \sum_{\langle i,j\rangle} ( a_i b_j +a_i^\dagger b_j^\dagger)=q~\xi_{\bf k}\l(\sum_{\bf k}a_{\bf k}b_{{\bf k}}+\sum_{\bf k}a^\dagger_{\bf k}b^\dagger_{{\bf k}}\r),
\end{equation}
where $\xi_{\bf k}=\dfrac{1}{q}\l(\sum_{\boldsymbol{\delta}}\e^{-i{\bf k}\cdot \boldsymbol{\delta}}\r)$ is assumed to be real, which is true for a centrosymmetric material. Using these transformations, the part of  Hamiltonian (\ref{Hreal}) containing only two magnon operators can be written in the ${\bf k}$-space as
\begin{equation}
    \mathcal{H}_{2}=E_a\sum_{\bf k}a_{\bf k}^\dagger a_{\bf k}+E_b\sum_{\bf k}b_{\bf k}^\dagger b_{\bf k}+ J S \xi_{\bf k} q\sum_{\bf k}\l(a_{\bf k}b_{{\bf k}}+a^\dagger_{\bf k}b^\dagger_{{\bf k}}\r),
\end{equation}
where $E_{a(b)}=[Jq+2K]S+(-)\gamma \mu_0 \hbar H_0$. The magnon eigenmodes are obtained by diagonalizing $\mathcal{H}_2$. Performing the Bogoliubov-Valatin transformations,
\begin{equation}
    a_{\bf k}=u_{\bf k}\alpha_{\bf k}-v_{\bf k}\beta^\dagger_{\bf k}, 
\end{equation}
\begin{equation}
    b_{\bf k}=-v_{\bf k}\alpha^\dagger_{\bf k}+u_{\bf k}\beta_{\bf k}, 
\end{equation}
$\mathcal{H}_2$ gets diagonalized as
\begin{equation}
    \mathcal{H}_2=\sum_{\bf k}\hbar \omega_\alpha({\bf k})\alpha_{\bf k}^\dagger \alpha_{\bf k}+\sum_{\bf k}\hbar \omega_\beta({\bf k})\beta_{\bf k}^\dagger \beta_{\bf k} + \text{constant},
\end{equation}
where $~\omega_{\alpha(\beta)}({\bf k})=\frac{S}{\hbar}\sqrt{J^2q^2(1-\xi^2_{\bf k})+4K(Jq+K)}+(-)\gamma\mu_0H_0 , ~u_{\bf k}=\cosh (\dfrac{\phi_{\bf k}}{2}),~ v_{\bf k} = \sinh(\dfrac{\phi_{\bf k}}{2}),~ \phi_{\bf k}=\text{arctanh} \l[\dfrac{\xi_{\bf k}}{(2\tilde{K}+1)}\r]$ and $\tilde{K}=\dfrac{K}{Jq}$.

Usually, only the two-magnon terms are retained to evaluate the magnon spectrum as the remaining nonlinear (four-magnon) terms are vanishingly small in the macroscopic limit $(N\to\infty)$. However, we must consider the nonlinear terms to obtain the required dispersive coupling. The Fourier transformation of the nonlinear terms in Eq. (\ref{Hreal}) gives 
\begin{equation}
    \sum_{\langle i,j\rangle}a_i^\dagger a_i  b_j^\dagger b_j =  \frac{q}{N}\sum_{{\bf k}, {\bf k}^\prime, {\bf Q} } \xi_{-\bf Q} ~a^\dagger_{\bf k}a_{\bf k-Q}b_{{\bf k}^\prime}^\dagger b_{{\bf k}^\prime-{\bf Q}} 
\end{equation}
\begin{equation}
    \sum_{\langle i,j\rangle}a_i   b_j^\dagger b_j b_j \text{ + H.c.}=  \frac{q}{N}\sum_{{\bf k}, {\bf k}^\prime, {\bf Q} } \xi_{\bf k} ~a_{\bf k}b_{\bf k^\prime+Q}^\dagger b_{{\bf k}^\prime} b_{{\bf k}+{\bf Q}} \text{ + H.c.}
\end{equation}
\begin{equation}
    \sum_{\langle i,j\rangle}a_i^\dagger   a_i a_i  b_j \text{ + H.c.}=  \frac{q}{N}\sum_{{\bf k}, {\bf k}^\prime, {\bf Q} } \xi_{\bf k^\prime} ~a^\dagger_{\bf k+Q}a_{\bf k^\prime+Q} a_{\bf k} b_{{\bf k}^\prime} \text{ + H.c.}
\end{equation}
\begin{equation}
    \sum_{i}(a_i^\dagger   a_i)^2 =  \frac{1}{N}\sum_{{\bf k}, {\bf k}^\prime }  ~a^\dagger_{\bf k^\prime+Q}a_{\bf k+Q}a_{{\bf k}}^\dagger a_{{\bf k}^\prime} 
\end{equation}
\begin{equation}
    \sum_{j}(b_j^\dagger   b_j)^2 =  \frac{1}{N}\sum_{{\bf k}, {\bf k}^\prime }  ~b^\dagger_{\bf k^\prime+Q}b_{\bf k+Q}b_{{\bf k}}^\dagger b_{{\bf k}^\prime}. 
\end{equation}

From the above equations, we see that there are terms that mix different ${\bf k}$'s. Rewriting them in terms of the eigenmode operators $\alpha_{\bf k}$ and $\beta_{\bf k}$, we extract the dispersive interaction $\sim\alpha^\dagger_{\bf k}\alpha_{\bf k} \beta_{\bf k^\prime}^\dagger\beta_{\bf k^\prime}$ out.  The terms which are off-diagonal in $\{|m\rangle_{\alpha,{\bf k}}|n\rangle_{\beta,{\bf k^\prime}}\}$ basis result in higher-order $(\sim1/N^2)$ corrections in the magnon frequencies and can be neglected in our analysis. Retaining the dispersive correction, the new magnon Hamiltonian can be expressed as
\begin{equation}
   \mathcal{H}=\sum_{\bf k}\hbar \omega_\alpha({\bf k})\alpha_{\bf k}^\dagger \alpha_{\bf k}+\sum_{\bf k}\hbar \omega_\beta({\bf k})\beta_{\bf k}^\dagger \beta_{\bf k} +\sum_{{\bf k,k^\prime}} \hbar \chi_{{\bf k,k^\prime}}\alpha^\dagger_{\bf k}\alpha_{\bf k} \beta_{\bf k^\prime}^\dagger\beta_{\bf k^\prime}.
\end{equation}
We refrain from providing the expression of $\chi_{\bf k,k^\prime}$ as it is long and cumbersome. The term essentially implies that there would be dispersive shifts in frequency of the mode $\beta_{\bf k^\prime}$, i.e. $\omega_\beta({\bf k^\prime})\to  \omega_\beta({\bf k^\prime})+\chi_{\bf k,k^\prime}~n_{\alpha,{\bf k}}$, upon occupation of the modes $\{\alpha_{\bf k}\}$ with different momenta and vise-versa.  
\begin{figure}[t]
\centering
\includegraphics[trim={0cm 0cm 0cm 0cm},clip,width=10cm]{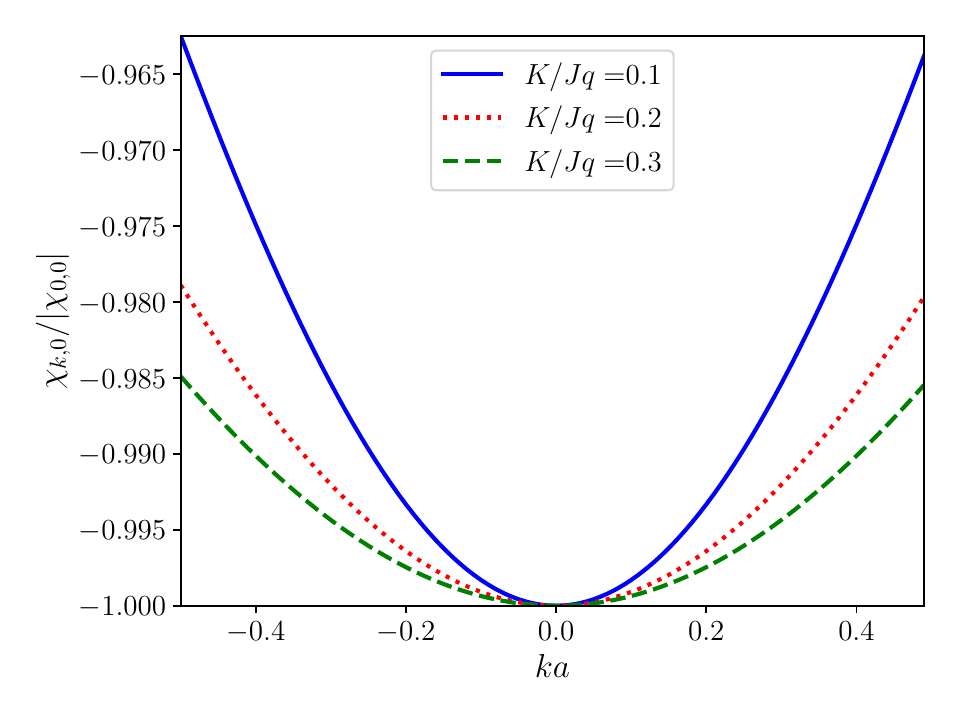}
\caption{Variation of strength of the dispersive interaction of mode $\beta_{\bf 0}$ with momenta of the $\alpha_{\bf k}$ modes for different values of anisotropy. As an example, we have considered a one-dimensional AFM with two sublattices so that $\xi_{\bf k}=\cos(ka/2)$, where $a$ is the length of the unit cell. We find that the dispersive interaction with $\alpha_{{\bf k}\neq0}$ modes is finite and will leave their signatures on the spectroscopy of $\beta_{\bf 0}$ mode, provided the former are sufficiently occupied.}
\label{disp}
\end{figure}

We are interested in using only the uniform mode $\beta_{{\bf k}=0}$ as the probe to sense the magnonic superpositions in the mode $\alpha_{{\bf k}=0}$. These are the modes that are typically excited in antiferromagnetic resonance experiments. A resonant spectroscopy of $\beta_{{\bf k}=0}$ mode resolves the magnon number states $\{n_{\alpha,0}\}$ of the $\alpha_{{\bf k}=0}$ mode through the dispersive shifts $\omega_\beta(0) +\chi_{0,0} n_{\alpha,0}$ in the former's frequency. However, the $\alpha_{{\bf k}\neq0}$ modes may also be populated due to thermal excitations, magnon-phonon or magnon-magnon scatterings, etc. They will impart additional dispersive shifts which may interfere with the sensing of the desired mode. The strength of dispersive interaction of the $\beta_{{\bf k}=0}$ mode  with a given $\alpha_{\bf k}$ mode is evaluated as
\begin{equation}
    \chi_{{\bf k},0}=\dfrac{Jq}{\hbar N}\l[\frac{1}{2}\l\{ \sinh(\phi_0+\phi_{\bf k})-\cosh{(\phi_{0}+\phi_{\bf k}})-1\r\}-2\tilde{K}(\cosh{\phi_{\bf k}}\cosh{\phi_{0}}-1) \r],
\end{equation}
where $\phi_0\equiv \phi_{{\bf k}=0}$. 
The strength of the dispersive coupling with $\alpha_{{\bf k}= 0}$ mode is $\chi_{0,0}=-\frac{Jq}{\hbar N}$.
The variation of $\chi_{{\bf k},0}$ with $k$ is shown in Fig \ref{disp}. The plot shows that the dispersive strength for a wide range of ${\bf k}$ around ${\bf k}=0$ is of the same order of magnitude as $|\chi_{0,0}|$. So, the spectroscopy of $\beta_{{\bf k}=0}$ mode can have significant contributions from all of them.  These contributions can be distinguished by making a smaller magnet. In that case, the magnon frequencies of the highly discretized ${\bf k}\neq0$ modes are separated from those of the lowest $\alpha_{{\bf k}=0}$ mode by several GHz \cite{Skogvoll2021}. So, the sets of mutivalued frequencies of $\beta_{{\bf k}=0}$ mode corresponding to the occupation of  the different $\alpha_{{\bf k}\neq0}$ modes will be well separated from each other. Based on these arguments, we retain only the contribution from the ${\bf k}=0$ mode.

%Since we would like to use However it shows that However, this mixing is not relevant for the physics pertaining to our work. This can be explained as follows. 
%We can hence discard the nonlinear interactions between the ${\bf k}\neq0$ modes as they are not relevant for the spectrocopy we are interested in. %Furthermore, Since the strength of the mixing terms $\propto 1/N$, they are a few orders of magnitude smaller than the magnon energies despite the smaller dimensions of the magnet, and do not affect the uniform modes significantly. The set of finite ${\bf k}$ modes can also be treated as a Markovian bath for the uniform modes on account of the weak interaction between the modes and quasi-continuous nature of ${\bf k}$. 

Defining $\alpha_{{\bf k}=0}\equiv\alpha$ and $\beta_{{\bf k}=0}\equiv \beta$, the magnonic Hamiltonian of these modes is derived as
\begin{equation}
    \mathcal{H}_{({\bf k}=0)}=\hbar \omega_\alpha \alpha^\dagger \alpha + \hbar \omega_\beta \beta^\dagger \beta -\frac{Jq}{N} \alpha^\dagger \alpha \beta^\dagger \beta  - \frac{K}{N} \l[(\alpha^{\dagger} \alpha)^2 + (\beta^{\dagger} \beta)^2\r]+\mathcal{H^\prime},
\end{equation}
where $u=\cosh (\frac{\phi}{2}),~ v = \sinh(\frac{\phi}{2})$, 
$\phi=\text{arctanh} \l[1/(2\tilde{K}+1)\r], ~\tilde{K}=\frac{K}{Jq}$.
The bare magnon frequencies are $\omega_{\alpha(\beta)}= \omega_0 + (-) \gamma \mu_0 H_0$, where $\omega_0=2S\sqrt{K(Jq+K)}/\hbar$. The term $-\frac{Jq}{N}\alpha^\dagger \alpha \beta^\dagger \beta$ is the direct dispersive interaction necessary to resolve the superpositions. It is to be noted that detuning between the two modes is not necessary to get this interaction, unlike the \textit{effective} dispersive interaction (discussed in Sec. \ref{effective-dispersive}). 

The terms $- \frac{K}{N} \l[(\alpha^{\dagger} \alpha)^2 + (\beta^{\dagger} \beta)^2\r]$ are Kerr  nonlinearities which affects the resolution at high temperatures (discussed in details in Sec. \ref{Kerr}). These effects can be minimized by choosing a very low anisotropy strength $K$. The third contribution $\mathcal{H}^\prime$ contains frequency-pulling terms ($\sim \alpha^\dagger \alpha$, $\sim \beta^\dagger \beta$) and several mixed terms in the magnon operators that are counter-rotating in the interaction picture. The frequency-pulling terms lead to a constant shift in the magnon frequencies  which are much smaller $\sim\mathcal{O}(1/N)$ than the bare frequencies. They can be neglected as they do not affect the resolution of the magnon number states. The counter-rotating terms can be discarded by choosing frequencies so that rotating wave approximation is valid. Hence, ignoring the Kerr-nonlinearities and $\mathcal{H}^\prime$, the magnon Hamiltonian of the AFM in terms of its eigenoperators can be expressed as
 \begin{equation}\label{dispersive}
     \mathcal{H}_0=\hbar \omega_\alpha \alpha^\dagger \alpha + \hbar \omega_\beta \beta^\dagger \beta -\frac{Jq}{ N}\alpha^\dagger \alpha \beta^\dagger \beta.
 \end{equation}

\section{Simulation of the spectroscopy}\label{sec: simulation}

In this section, we describe the simulation of the spectroscopy in detail. 
\subsection{Formalism}\label{subsec: formalism}
The magnon mode $\alpha$ is probed by the bosonic mode $\beta$. Let the magnonic state be $|\psi\rangle=\sum_{n_\alpha}^{}c_{n_\alpha} |n_\alpha \rangle$. The probability distribution $\l\{|c_{n_\alpha}^2|\r\}=\l\{P_{n_\alpha}\r\}$ is to be calculated using the spectroscopy. The probe is connected to a thermal bath at temperature $T$ and subjected to a coherent drive $\mathcal{V}(t)= d\cos\omega t (\beta + \beta^\dagger)$ of strength $d$.
The Hamiltonian of the magnon-probe system is 
\begin{equation}
    \mathcal{H}(t)=\hbar \omega_\alpha \alpha^\dagger \alpha + \hbar \omega_\beta \beta^\dagger \beta + \hbar \chi \alpha^\dagger \alpha \beta^\dagger \beta + \mathcal{V}(t).
\end{equation}
where $\chi$ is the dispersive strength for a generic bosonic probe. For the AFM magnon mode as probe, $\chi=-Jq/\hbar N<0$. The time evolution is studied using the Lindblad master equation \cite{Breuer2002, Lindblad1976}:
\begin{equation}\label{Lindblad}
    \frac{d\rho}{dt}=-\frac{i}{\hbar}[\mathcal{H}(t),\rho(t)] + \sum_{n=1,2}^{} C_n \rho(t) C_n^\dagger -\frac{1}{2}\l\{\rho(t), C_n^\dagger C_n\r\}
\end{equation}
where $C_n$ are the collapse operators accounting for interaction with the bath. For the thermal bath, the collapse operators that characterize emission and absorption are $C_1=\sqrt{\kappa(n^\beta_\text{th}+1)} ~\beta$ and $C_2=\sqrt{\kappa~n^\beta_\text{th}}~ \beta ^\dagger$ respectively. Here, $\kappa$ is the decay rate and $n^\beta_\text{th}=[e^{\hbar \omega_\beta/k_B T}-1]^{-1}$ is the mean occupation number of the probe if it were in thermal equilibrium with the bath.

\subsection{Simulation}\label{subsec: simulation}

As an example, let the magnon mode be a initialized in a coherent state:
\begin{equation}\label{coh}
    |\psi\rangle = \e^{-\frac{|\alpha|^2}{2}}\sum_{n}^{N_\alpha}\frac{\alpha^{n_\alpha}}{\sqrt{n_\alpha!}}|n_\alpha\rangle.
\end{equation}
The density matrix of the magnon mode is $\rho_\alpha=|\psi\rangle \langle \psi|$. The probe mode can be initialized in any arbitrary state $|\psi\rangle_\beta$ for which the density matrix would be $\rho_\beta=|\psi\rangle_\beta \langle \psi|_\beta$. So, the initial density matrix of the total system is  $\rho{(0)}=\rho_\alpha\otimes \rho_\beta.$  
For simplicity, the probe is initialized in vacuum $|0\rangle$. The time-evolved $\rho_{}(t)$ is obtained by solving the master equation (\ref{Lindblad}) numerically. The mean excitation or boson number of the probe is obtained as $\langle \beta^\dagger \beta \rangle= \text{Tr}[\rho (t)~\beta^\dagger \beta]$. \\

\textit{Parameters used:} We consider a truncated Hilbert space $N_\alpha=6$ as the Hilbert space dimension of the magnon mode does not affect the simulation results. Upto moderately high temperatures $k_BT\sim \mathcal{O}(\hbar\omega_\beta)$ and  driving strengths $d/\hbar\omega_\beta\sim \mathcal{O}(10^{-2})$, taking $N_\beta=10$ is enough to produce qualitatively correct results and save computational time. However, higher $N_\beta$ would give more accurate results but at the expense of a much longer computational time. The following parameters are constant for all simulations: $\alpha=1.0, \chi/\omega_\beta=0.1,  \kappa/\omega_\beta=0.003$. For the case of weak driving and practically zero temperature, we choose $d/\hbar\omega_\beta=0.001$ and $k_BT/\hbar\omega_\beta =0.01$. The simulation results are independent of the values of $\omega_\alpha$ and $\omega_\beta$. \\

\textit{Observations:}
 \begin{enumerate}
     \item For a given frequency, the probe excitation reaches a steady value after a time around $\sim 15
     /\kappa$ with small fluctuations about it. A time-averaged $\langle \beta^\dagger \beta \rangle$ is taken around $t=20/\kappa$.
     \item A plot of steady state  $\langle \beta^\dagger \beta \rangle$ vs. $\omega$  shows peaks at $\omega=\omega_\text{res}^{n_\alpha}=\omega_\beta + n_\alpha \chi$ where $n_\alpha=0,1,2,...N_\alpha$. The shift between the resonant peaks is $|\chi|$ and the width is $\sim \kappa$. For proper resolution of the peaks, we choose $\kappa\ll |\chi|$.
     \item The value of $\langle \beta^\dagger \beta\rangle$ away from the resonances is equal to mean thermal occupation $n^\beta_\text{th}$ and we call it the baseline. The baseline is $\approx 0$ at $k_BT/\hbar \omega_\beta=0.01$, as expected.
     \end{enumerate}

\subsection{Resolving the magnon number states}
\label{subsec: resolving}

% \item {\bf Resolution of the peaks: }The peaks can be resolved well if $\kappa \ll \chi$. This is ensured in our simulation since we choose $\kappa_\beta=0.003~\hbar\omega_\beta$ and $\chi=0.1 ~\hbar\omega_\beta$.

    The relation between the probe excitation $\langle \beta^\dagger \beta \rangle_{n_\alpha}$ at a resonance $\omega=\omega^{n_\alpha}_\text{res}$ (corresponding to tip of the peak) and the probability distribution of the magnon mode $P_{n_\alpha}$ can be written as 
\begin{equation}\label{empirical}
    \langle \beta^\dagger \beta \rangle_{n_\alpha}= k P_{n_\alpha} + n^\beta_\text{th}, 
\end{equation}
where $k $ is a constant of proportionality that depends on the driving strength and decay rate. The analytical derivation of Eq. (\ref{empirical}) is shown in Sec. \ref{lindblad-section}.  Equation (\ref{empirical}) implies that the net excitation is an algebraic sum of driving and thermal contributions \cite{Drummond_1980,Gardiner2004}. The part $kP_{n_\alpha}=\langle \beta^\dagger \beta \rangle_{n_\alpha}-n^\beta_\text{th} $ is the peak height measured from the baseline. The above equation can be verified by calculating $k$ from one of the resonances, say at $\omega_\text{res}^{n^\prime_a}$, for the known state of the magnon mode [Eq. (\ref{coh})]. That is,
\begin{equation}
    k=\frac{1}{P_{n^\prime_\alpha}}\l[\langle \beta^\dagger \beta \rangle_{n^\prime_\alpha}-n^\beta_\text{th}  \r],
\end{equation}
where $P_{n^\prime_\alpha}$ is known and $\langle \beta^\dagger \beta \rangle_{n^\prime_\alpha}$ is obtained from the simulation. Using this value of $k$, we can calculate the other $P_{n_\alpha}$ as
\begin{equation}\label{Pna}
    P_{n_\alpha}=\frac{1}{k}\l[\langle \beta^\dagger \beta \rangle_{n_\alpha}-n^\beta_\text{th}  \r],
\end{equation}
%=P_{n_\alpha^\prime}\l(\frac{\langle \beta^\dagger \beta \rangle_{n_\alpha}-n^\beta_\text{th}}{\langle \beta^\dagger \beta \rangle_{n_\alpha^\prime}-n^\beta_\text{th}}\r)
which are in good agreement with the known values of $P_{n_\alpha}$ considered in the simulation.

 For experimental measurements, all the values of $P_{n_\alpha}$ are unknown and should be measured from the spectroscopy. So, the approach of calculating $P_{n_\alpha}$ from one of the known values fails. In this case, $k$ can be calculated using the normalization condition: $\text{Tr}(\rho_\alpha)=\sum_{m_\alpha}^{} P_{m_\alpha}=1$. Using Eq. (\ref{empirical}), we can write
\begin{equation}
    k=\sum_{m_\alpha}\l[\langle \beta^\dagger \beta \rangle_{m_\alpha}-n^\beta_\text{th}  \r].
\end{equation}
This means that the proportionality constant is equal to the sum of the peak heights measured from the baseline. Using this $k$ in Eq. (\ref{Pna}), we get the desired $\{P_{n_\alpha}\}$ of the probed bosonic mode as
\begin{equation}
    P_{n_\alpha}=\frac{\langle \beta^\dagger \beta \rangle_{n_\alpha}-n^\beta_\text{th} }{\sum_{m_\alpha}\l[\langle \beta^\dagger \beta \rangle_{m_\alpha}-n^\beta_\text{th} \r]}~.
\end{equation}

% {\color{blue}\item {\bf Dependence on driving strength:} We predict that $\alpha$ is proportional to $d^2$ and inversely proportional to $\kappa^2$.}

    The occupation probabilities $\{P_{n_\alpha}\}$ in the steady state are identical to those in the initial state of the magnon mode. %This can be checked by matching $\{P_{n_\alpha}\}$ with the diagonal elements of the reduced density matrix $\text{Tr}_\beta\l(\rho \r)$ in the Fock basis. 
    This shows that the nature of measurement is Quantum Non-Demolition type \cite{Milburn_1983,Imoto_1985,Munro_2005,Helmer_2009} where no magnon is destroyed or created.

\section{Comparison between qubit and bosonic mode as sensors of magnon number state}\label{sec: comparison}

\begin{figure}[htbp]
		\centering
\includegraphics[trim={0cm 0cm 0cm 0cm},clip,width=13cm]{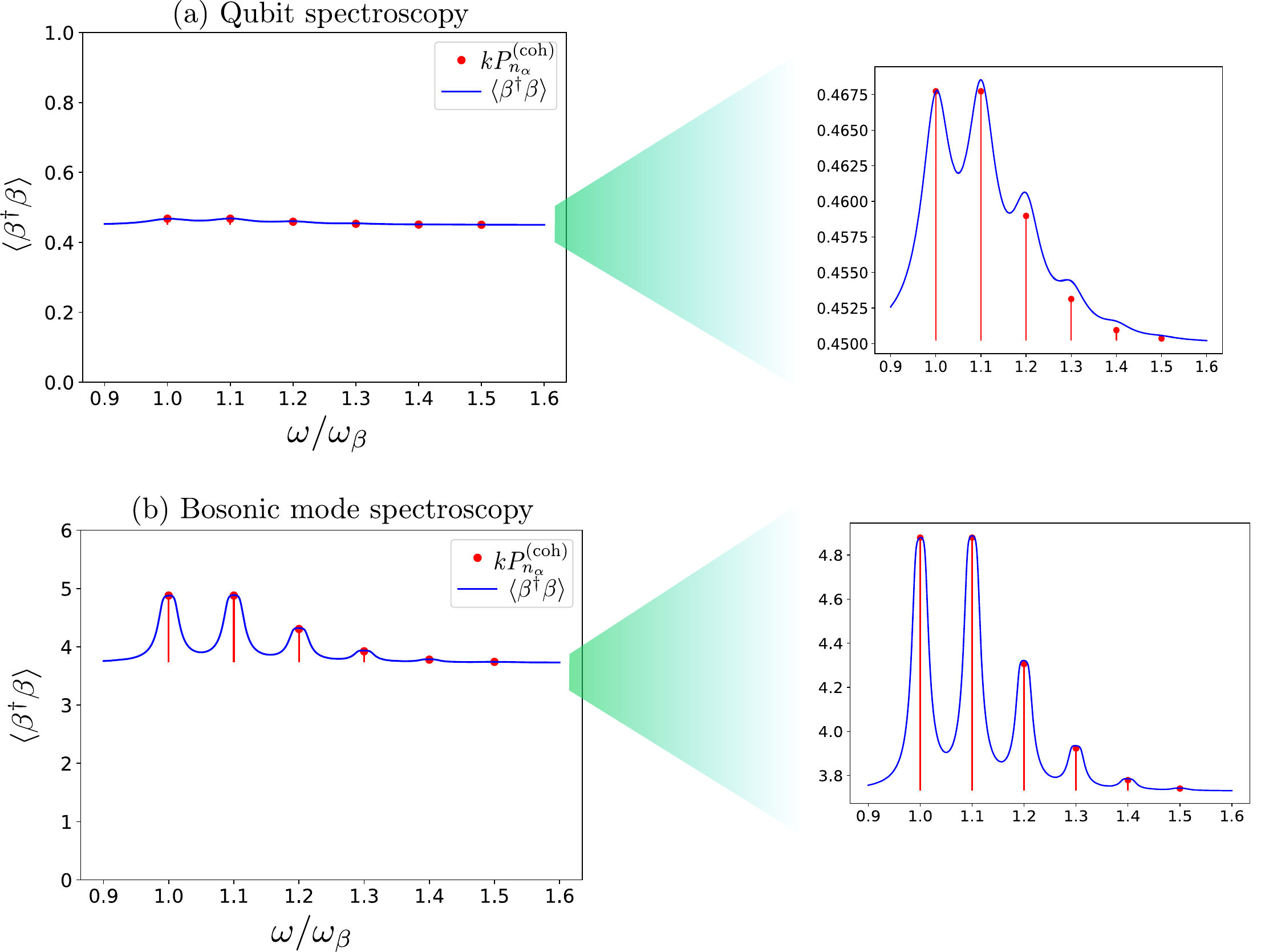}
		\caption{Simulation data for a coherent magnon state using (a) qubit and (b) bosonic mode as a probe, at high temperature and strong drive. The bosonic probe can still resolve the magnonic number states at higher temperature and stronger drive, while the qubit cannot. This is attributed to the infinitely large Hilbert space of the bosonic mode which does not put a constraint on its maximum excitation, unlike the case of a qubit.}
		\label{qubit}
	\end{figure}
% \subsection{Conclusions}
%  \begin{enumerate}
%      \item The resonant peaks do not get distorted even at stronger driving or higher temperatures.
%      \item  At higher temperatures, the baseline simply gets raised due to thermal excitation. Then, the peaks are very small as compared to the baseline and hence difficult to resolve. In that case, a stronger drive enhances the visibility of the peaks by increasing their heights, just as in a classical resonance.
%      \item Low driving amplitudes suffice when $T\to 0$ as $n^\beta_\text{th}(T) \approx 0$ at thermal equilibrium.
%      \item The steady state is independent of the initial state of the probe mode. This is advantageous for the experimental measurement as the probe need not be initialized in the ground state or any other fixed state for that matter. So, measurements can be performed sequentially without having to reinitialize the system.
%  \end{enumerate}

In this section, we highlight the difference between the performances of a qubit and a bosonic mode as probes at high temperature and strong driving. The Hamiltonian of a qubit dispersively coupled with the magnon mode is expressed as \cite{Savage1990, Faria1999, Blais2004,Zueco2009}
\begin{equation}
    \mathcal{H}_\text{qm}=\hbar \omega_\alpha \alpha^\dagger \alpha + \frac{\hbar\omega_q}{2} \sigma_z + \frac{\hbar\chi}{2} \alpha^\dagger \alpha \sigma_z. \end{equation}
where $\omega_q$ is the bare qubit frequency. Due to the dispersive coupling, the qubit frequency becomes multivalued as $\omega_q+n_\alpha \chi$. Using the same spectroscopic setup as for the bosonic probe, we obtain a plot of steady-state qubit excitation $\langle \sigma_+ \sigma_-\rangle$ as a function of $\omega$ on being subjected to a drive $\mathcal{V}_q(t)=d \cos \omega t (\sigma_+ + \sigma_-).$ 

The simulation data for high temperature $k_BT/\hbar\omega_{q(\beta)}=5.0$ and strong driving $d/\hbar\omega_{q(\beta)}=0.05$ is shown in Fig. \ref{qubit} a(b) for the qubit (bosonic mode) as probe. We find that the peaks are highly diminished for the case of a qubit even under strong drives, but still well resolved for the bosonic probe. The zoomed plots on the right panel show that the resolvability of the peaks in the qubit spectroscopy is low as the adjacent peaks tend to overlap and their proportionality with the exact magnon probability distribution is hampered. The bosonic mode spectroscopy still shows a good resolution of the peaks keeping their proportionality with the probability distribution intact. In the simulation, we consider $N_\beta=15$. The heights (linewidths) of the peaks for the bosonic probe can be further increased (decreased) by choosing a larger $N_\beta$ in the simulation which gives an improved resolution but requires more computational time. This shows that the unbounded Hilbert space of the bosonic mode makes it a better sensor of number states than a qubit under nonideal conditions such as high temperatures and strong drives.

\section{Effective dispersive interaction between a magnon and phonon mode}\label{effective-dispersive}

In this section, we show that an effective dispersive interaction can also be engineered between a magnon and phonon mode from their nonlinear couplings. The magnon-phonon Hamiltonian can be expressed as
\begin{equation}
    \mathcal{H}_\text{mp}=\mathcal{H}_{\alpha\beta} + V. 
\end{equation}
where $\mathcal{H}_{\alpha\beta}=\hbar \omega_\alpha \alpha^\dagger \alpha + \hbar \omega_\beta \beta^\dagger \beta$, with $\alpha$ and $\beta$ as the magnon and phonon annihilation operators respectively, and $V\equiv V(\alpha,\beta,\alpha^\dagger,\beta^\dagger)$ being a generic coupling between the two modes. When the frequencies of the modes are highly detuned with respect to each other, say $\omega_\beta\gg \omega_\alpha$, a direct exchange of energy between the modes is highly suppressed and virtual absorption-emission processes dominate. An effective Hamiltonian that takes into account these processes can be derived using a Schrieffer-Wolff transformation (SWT) \cite{Schrieffer1966,Landi2024}. 

The magnon-phonon coupling $V$ can be decomposed as $V=\sum_{\Omega>0} V_\Omega+V_\Omega^\dagger$ where $V_\Omega$ is an eigenoperator of $\mathcal{H}_{\alpha \beta}$, i.e. $[\mathcal{H}_{\alpha \beta},V_{\Omega}]=-\hbar \Omega V_{\Omega}$. 
Using SWT and retaining processes upto second order in the coupling strength(s), the effective Hamiltonian for the subspace of $\alpha$ mode is expressed as \cite{Landi2024}
\begin{equation}\label{SWT}
    [\mathcal{H}_\text{mp}]_\text{eff}=\mathcal{H}_{\alpha \beta} -\frac{1}{2\hbar}{\sum_{\Omega,\Omega^\prime>0}}^\prime\l(\frac{1}{\Omega} + \frac{1}{\Omega^\prime} \r)[V_{\Omega},V_{\Omega^\prime}^\dagger].
\end{equation}
 The prime in the summation means that only those pairs of eigenoperators $(V_\Omega, V_{\Omega^\prime}^\dagger)$ should be chosen for which $|\Omega-\Omega^\prime|\ll \omega_\beta$, considering the far-detuned limit $\omega_\beta\gg \omega_\alpha$. This ensures virtual transitions back to the subspace of $\alpha$ mode. 
 
 The magnon-phonon couplings \cite{Abrahams1952} are usually linear $\sim\alpha^\dagger \beta$ + H.c. \cite{Potts2021} or quadratic $\propto \alpha^{\dagger2} \beta, \alpha^{\dagger2} \beta, \alpha^\dagger\alpha \beta $ + H.c. \cite{Rueckriegel2014, Ballestero2020_prb, Bittencourt2023,kuenstle2025} in the magnon operators. 
 For a linear coupling $V=\hbar\zeta_1 (\alpha^\dagger \beta+ \alpha \beta^\dagger) + \hbar \zeta_2(\alpha\beta+\alpha^\dagger \beta^\dagger)$, the eigenoperators $\{V_\Omega\}$ with $\Omega>0$ are $V_{\omega_\beta-\omega_\alpha}=\hbar \zeta_1 \alpha^\dagger \beta$ and $V_{\omega_\beta+\omega_\alpha}=\hbar \zeta_2 \alpha \beta$. 
 Then, the effective Hamiltonian obtained using (\ref{SWT}) is 
\begin{equation}
    \l[\mathcal{H}_\text{mp}\r]_\text{eff}^l= \mathcal{H}_{\alpha\beta} -\frac{\hbar \zeta_1^2}{\omega_\beta-\omega_\alpha}(\alpha^\dagger \alpha-\beta^\dagger \beta)-\frac{\hbar \zeta_2^2}{\omega_\beta+\omega_\alpha}(\alpha^\dagger \alpha+\beta^\dagger \beta+1)-\frac{\hbar \zeta_1 \zeta_2 \omega_\beta}{\omega_\beta^2-\omega_\alpha^2}(\alpha^2+\alpha^{\dagger 2}).
\end{equation}
 Thus, we see that the linear coupling does not give rise to a dispersive interaction $\sim \alpha^\dagger \alpha \beta^\dagger \beta$ in the effective Hamiltonian. The parametric magnon-phonon coupling $V=\hbar \eta ~\alpha^\dagger \alpha(\beta^\dagger+\beta)$, which is usually considered the most relevant in far-detuned limit \cite{Zhang2016,Li2018,Li2019,Ballestero2020_prl} also fails to produce a dispersive interaction in its effective Hamiltonian $\l[\mathcal{H}_\text{mp}\r]^p_\text{eff}=\mathcal{H}_{\alpha\beta} -\frac{\hbar \eta^2}{\omega_\beta}\alpha^\dagger \alpha.$

Now, we show that the parametric-conversion terms $\sim\alpha^{\dagger2} \beta $ + H.c. as well as the counter-rotating terms $\sim \alpha^{2} \beta $ + H.c. give rise to the required dispersive interaction in the far-detuned regime. 
To establish design principles, we consider the most general three-particle interaction: 
\begin{equation} \label{V}
    V= \hbar \chi_1 (\alpha^\dagger \beta^2+\alpha \beta^{\dagger2}) + \hbar \chi_2 (\alpha \beta^2+\alpha^\dagger \beta^{\dagger2}) + \hbar \chi_3 (\alpha^2 \beta^\dagger+\alpha^{\dagger2} \beta) + \hbar \chi_4 (\alpha^2 \beta +\alpha^{\dagger2} \beta^\dagger).
\end{equation}
The terms with coefficients $\hbar \chi_2$ and $\hbar \chi_4$ are counter-rotating in the interaction picture. The eigenoperators for this $V$ are $V_{2\omega_\beta-\omega_\alpha}=\hbar \chi_1 \alpha^\dagger \beta^2,~V_{2\omega_\beta+\omega_\alpha}=\hbar \chi_2 \alpha \beta^2,~V_{\omega_\beta-2\omega_\alpha}=\hbar \chi_3 \alpha^{\dagger2} \beta$ and $V_{\omega_\beta+2\omega_\alpha}=\hbar \chi_4 \alpha^2 \beta$.
Using (\ref{SWT}), the effective Hamiltonian for the $V$  in Eq. (\ref{V}) can be written as
\begin{equation}
\begin{aligned}
    &[\mathcal{H}_\text{mp}]_\text{eff}=\mathcal{H}_{\alpha \beta} - \frac{\hbar \chi_1^2}{2\omega_\beta-\omega_\alpha}[\alpha^\dagger \beta^2, \alpha \beta^{\dagger2}] - \frac{\hbar \chi_2^2}{2\omega_\beta+\omega_\alpha}[\alpha \beta^2, \alpha^\dagger \beta^{\dagger2}] - \frac{\hbar \chi_3^2}{\omega_\beta-2\omega_\alpha}[\alpha^{\dagger2} \beta, \alpha^2 \beta^{\dagger}] -\frac{\hbar \chi_4^2}{\omega_\beta+2\omega_\alpha}[\alpha^2 \beta, \alpha^{\dagger2} \beta^\dagger ]\\
    & -\frac{\hbar \chi_1 \chi_2}{2} \l(\frac{1}{2\omega_\beta-\omega_\alpha}+\frac{1}{2\omega_\beta+\omega_\alpha}\r)\l([\alpha^\dagger \beta^2,\alpha^\dagger \beta^{\dagger2}]+ \text{H.c.}\r) -\frac{\hbar \chi_3 \chi_4}{2} \l(\frac{1}{\omega_\beta-2\omega_\alpha}+\frac{1}{\omega_\beta+2\omega_\alpha}\r)\l([\alpha^{\dagger2} \beta,\alpha^{\dagger2} \beta^{\dagger}]+ \text{H.c.}\r).
\end{aligned}
\end{equation}
 Upon evaluating the commutators, the above equation reduces to
\begin{equation}
\begin{aligned}
    [\mathcal{H}_\text{mp}]_\text{eff}=\mathcal{H}_{\alpha\beta} &-\frac{\hbar \chi_1^2}{2\omega_\beta-\omega_\alpha}\l(2\alpha^\dagger\alpha + 4 \alpha^\dagger \alpha \beta^\dagger \beta - \beta^{\dagger2} \beta^2\r)
    -\frac{\hbar \chi_2^2}{2\omega_\beta + \omega_\alpha}\l[2(1+\alpha^\dagger \alpha + 2\beta^\dagger \beta + 2\alpha^\dagger \alpha \beta^\dagger \beta)+\beta^{\dagger 2}\beta^2\r]\\&-\frac{\hbar \chi_3^2}{\omega_\beta-2\omega_\alpha}\l(\alpha^{\dagger2} \alpha^2-2\beta^\dagger \beta - 4 \alpha^\dagger \alpha \beta^\dagger \beta \r)- \frac{\hbar \chi_4^2}{\omega_\beta+2\omega_\alpha}\l(\alpha^2\alpha^{\dagger2} + 2 \beta^\dagger\beta + 4 \alpha^\dagger\alpha \beta^\dagger \beta\r)\\&-\frac{2\hbar \chi_1 \chi_2 \omega_\beta}{4\omega_\beta^2-\omega_\alpha^2}\l( 2\alpha^{\dagger2} + 4\alpha^{\dagger 2} \beta^\dagger \beta + 2 \alpha^2 + 4 \alpha^2 \beta^\dagger \beta \r)-\frac{\hbar \chi_3 \chi_4 \omega_\beta}{\omega_\beta^2-4\omega_\alpha^2}\l(\alpha^4+\alpha^{\dagger4}\r) .
\end{aligned}
\end{equation}
The effective Hamiltonian can be rewritten as
\begin{equation}
   [\mathcal{H}_\text{mp}]_\text{eff}= \mathcal{H}_{\alpha \beta} + \hbar \chi_\text{eff} ~\alpha^\dagger \alpha~ \beta^\dagger \beta + \mathcal{H}_1
\end{equation}
where
\begin{equation}
    \chi_\text{eff}=-\frac{4 \chi_1^2}{2\omega_\beta-\omega_\alpha}-\frac{4 \chi_2^2}{2\omega_\beta+\omega_\alpha}+\frac{4  \chi_3^2}{\omega_\beta-2\omega_\alpha}-\frac{4 \chi_4^2}{\omega_\beta+2\omega_\alpha}
\end{equation}
is the strength of the effective dispersive coupling and $\mathcal{H}_1$ contains frequency-pulling terms ($\sim \alpha^\dagger \alpha, \beta^\dagger \beta$), Kerr terms ($\sim \alpha^{\dagger^2} \alpha^2, \beta^{\dagger2} \beta^2$) etc. 
For the magnon-phonon system, such an interaction arises only in the far-detuned limit. The specific conditions which relate the relevant detunings with the nonlinear coupling strengths for the emergence of the dispersive interaction $\sim \alpha^\dagger \alpha \beta^\dagger \beta$ are \cite{Ding_2017}:  $|\chi_1|\ll|2\omega_\beta-\omega_\alpha|, |\chi_2|\ll|2\omega_\beta+\omega_\alpha|, |\chi_3|\ll|\omega_\beta-2\omega_\alpha|$ and $ |\chi_4|\ll|\omega_\beta+2\omega_\alpha|$.

The frequency pulling terms only cause constant global frequency shifts which do not affect the sensing protocol. The Kerr nonlinearity appears in the probe mode ($\beta$) due to the coupling terms with coefficients $\chi_1$ and $\chi_2$, but appears in the magnon mode ($\alpha$) due to the terms with coefficients $\chi_3$ and $\chi_4$. The appearance of a Kerr nonlinearity in the magnon mode does not affect the frequency of the probe mode and is hence not a problem for the resolving the magnon number states. Since Kerr nonlinearity in the probe can be detrimental to the probe's performance at finite temperatures (discussed in the next section), it is desirable to suppress this effect. Choosing a suitable operating point in the frequencies can reduce the Kerr nonlinearity.  For example, if we choose $\omega_\beta\gtrsim2\omega_\alpha$ (within the limit $|\chi_3|\ll|\omega_\beta-2\omega_\alpha|$), it enhances the term $ 4\chi_3^2/(\omega_\beta-2\omega_\alpha)$ in $\chi_\text{eff}$ but diminishes the contribution from the other terms. The dispersive strength also has contributions from counter-rotating terms ($\propto \chi_2^2, \chi_4^2$) indicating that the effective Hamiltonian is valid beyond the rotating wave approximation \cite{Zueco2009}. 

\begin{figure}[t]
		\centering
\includegraphics[trim={0cm 0cm 0cm 0cm},clip,width=13cm]{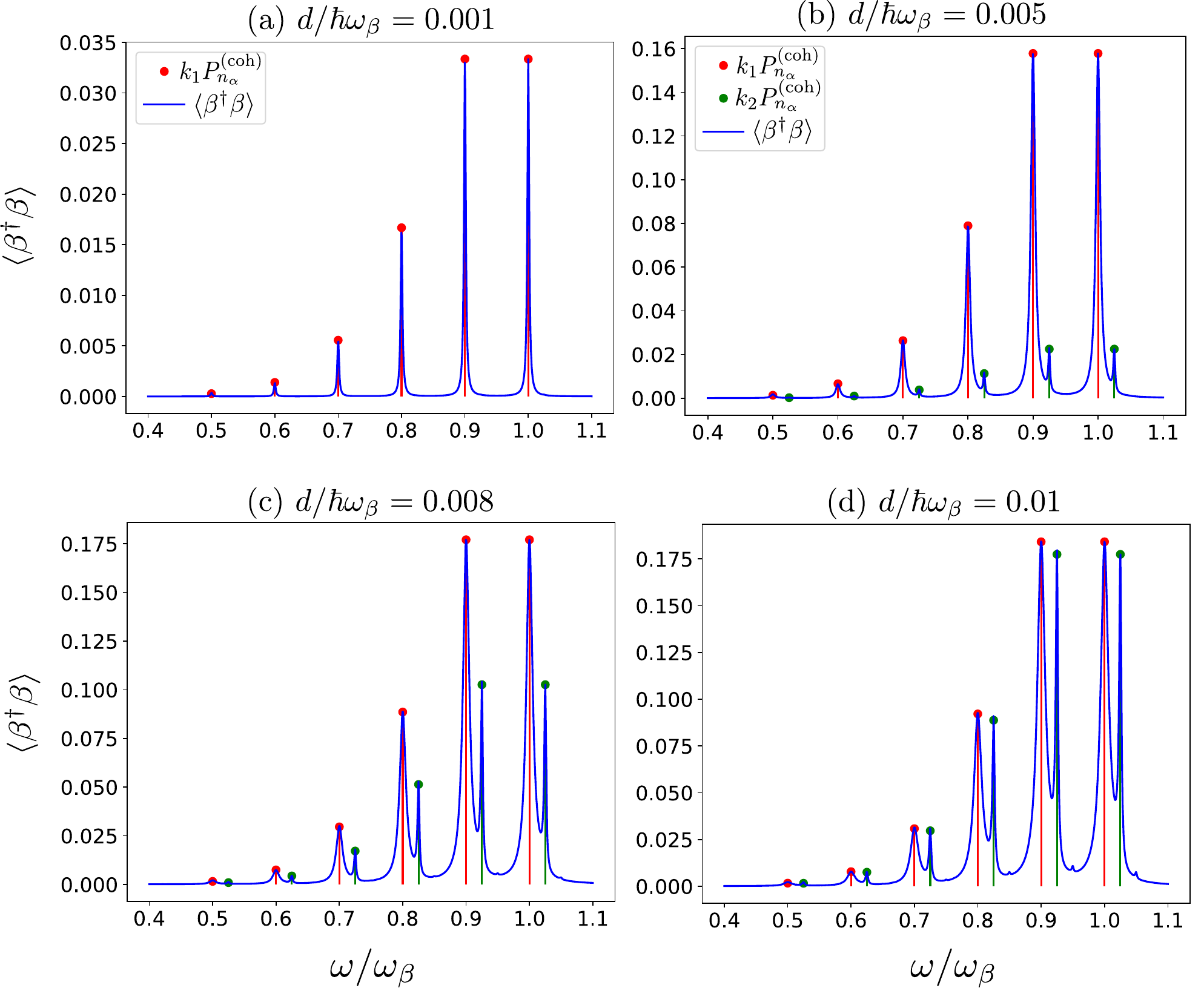}
		\caption{Simulation data of the probe bosonic mode for different driving amplitudes at a very low temperature ($k_BT/\hbar\omega_\beta=0.01$) in the presence of a Kerr nonlinearity. The new set of peaks (green dots) adjacent to the fundamental ones (red dots) arise due to two-photon absorption for the transition $|n_\alpha\rangle|0\rangle\to |n_\alpha \rangle |2\rangle$. Both sets of peaks are proportional to the probability distribution of the magnon mode. So, the bosonic mode can act as a sensor despite the Kerr nonlinearity if $k_BT\ll\hbar\omega_\beta$. }
		\label{lowT2}
	\end{figure}

\begin{figure}[h]
		\centering
\includegraphics[trim={0cm 0cm 0cm 0cm},clip,width=13cm]{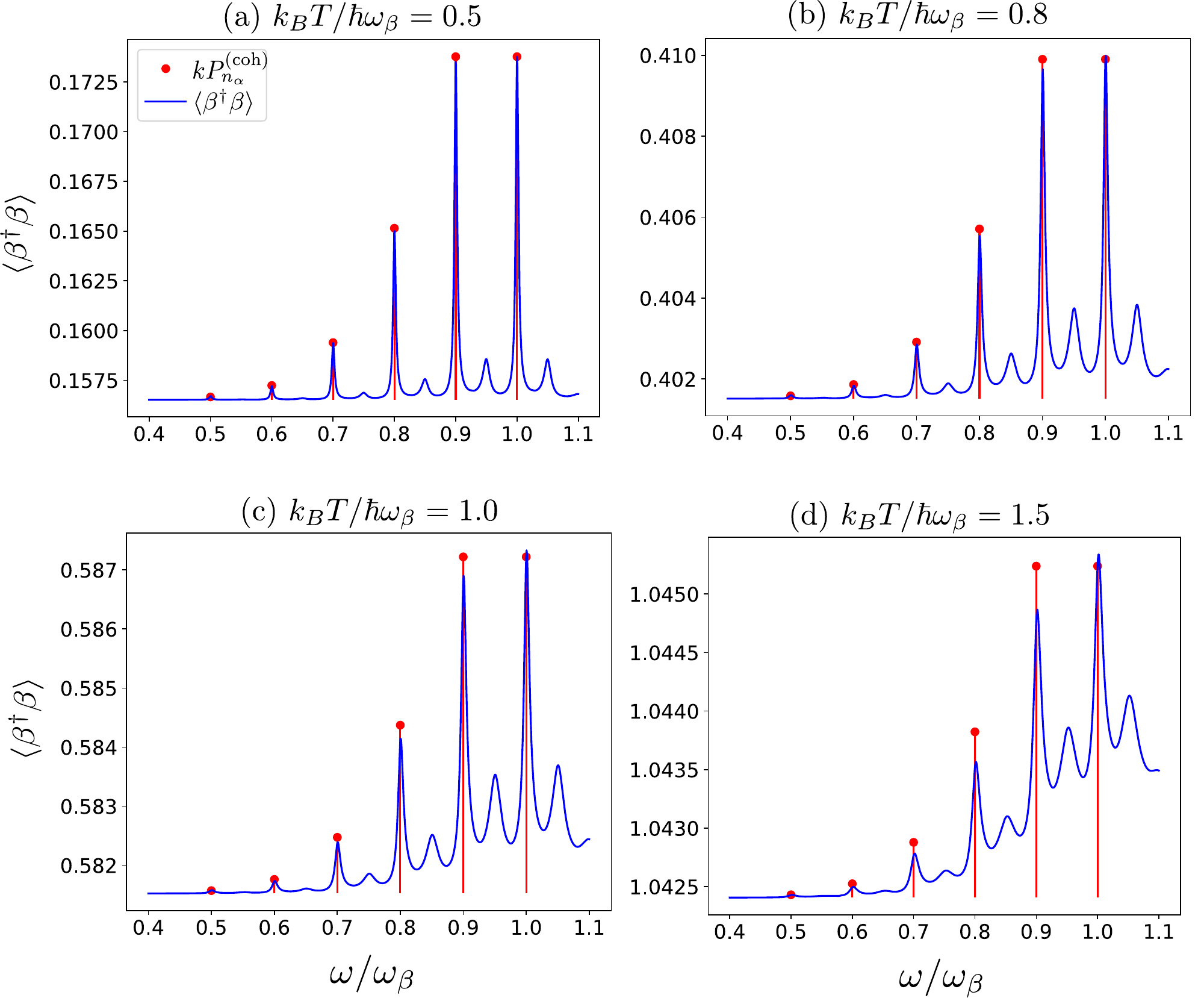}
		\caption{Simulation data for the probe bosonic mode at different temperatures and weak driving ($d/\hbar\omega_\beta=0.01$) in the presence of a Kerr nonlinearity. Additional peaks arise due to significant thermal occupation of excited levels $|n_\beta\rangle$ of the probe from which $|n_\alpha\rangle |n_\beta\rangle\to |n_\alpha\rangle |n_\beta+1\rangle$ transitions take place. These peaks overlap with the fundamental ones at $\omega_\text{res,1}^{n_\alpha}$ and alters their proportionality with the magnon number statistics.}
		\label{diff-T-d001}
	\end{figure}

\section{Effect of a Kerr nonlinearity in the probe bosonic mode}\label{Kerr}

In this section, we show how a Kerr nonlinearity can affect the resolvability of the magnon number states at high temperature. Let us consider only the effective Hamiltonian resulting from the first term in equation (\ref{V}):

\begin{equation}
   \mathcal{H}_\text{eff}=\hbar\omega_\alpha \alpha^\dagger \alpha  +\hbar\omega_\beta \beta^\dagger \beta - \frac{\hbar \chi_1^2}{2\omega_\beta-\omega_\alpha}\l(2\alpha^\dagger\alpha + 4 \alpha^\dagger \alpha \beta^\dagger \beta - \beta^{\dagger2} \beta^2\r).
\end{equation}
Here, $\beta^{\dagger 2} \beta^2$ is the Kerr term \cite{Gerry2005}.
%The off-diagonal terms in the Hamiltonian couple the Fock states with an energy difference of $
We define $\Delta\equiv2\omega_\beta-\omega_\alpha$. For the far-detuned case $\omega_\beta \gg \omega_\alpha$, $\Delta\equiv2\omega_\beta-\omega_\alpha$ is also large. %When $g \ll \Delta$, a Schrieffer-Wolf transformation gives us an effective Hamiltonian diagonal in the Fock-basis as
% \begin{equation}\label{SW1}
%     \mathcal{H}1^\text{SW}= \hbar \omega_\alpha \alpha^\dagger \alpha + \hbar \omega_\beta \beta^\dagger \beta + \frac{\hbar \chi_1^2}{\Delta}\l(  b^{\dagger 2} b^2  - 4 \alpha^\dagger \alpha \beta^\dagger \beta - 2 \alpha^\dagger \alpha \r)
% \end{equation}
% The effective interaction within the parenthesis has Kerr term ($b^{\dagger 2} b^2$), cross-Kerr term ($-4 \alpha^\dagger \alpha \beta^\dagger \beta$) and frequency-pulling term ($- 2 \alpha^\dagger \alpha$). The presence of the cross-Kerr term in the dispersive limit allows for usage of this system for a QND measurement. In this analysis, the cross-Kerr strength $\chi=-4 \chi_1^2/\Delta$. Due to negative sign of $\chi$, the resonances corresponding to higher values of $n_\alpha$ will be red-shifted with respect to $n_\alpha=0$. 
The Hamiltonian can be rewritten as
\begin{equation}
    \mathcal{H}_\text{eff}=\hbar\l(  \omega_\alpha - \frac{2 \chi_1^2}{\Delta} \r) \alpha^\dagger \alpha + \hbar\l[\omega_\beta+\frac{\chi_1^2}{\Delta}\l(-4\alpha^\dagger \alpha + \beta^\dagger \beta-1\r)\r]\beta^\dagger \beta.
\end{equation}
The energy levels are given by
\begin{equation}
    E_{n_\alpha,n_\beta}=\hbar\l(  \omega_\alpha - \frac{2 \chi_1^2}{\Delta} \r) n_\alpha + \hbar\l[\omega_\beta+\frac{\chi_1^2}{\Delta}\l(-4n_\alpha + n_\beta-1\r)\r]n_\beta.
\end{equation}

\[
\begin{array}{|c|c|}
\hline
\text{Energy Level} & \text{Expression} \\
\hline
E_{0,0} & 0 \\
\hline
E_{0,1} & \hbar\omega_\beta \\
\hline
E_{0,2} & 2\hbar \left(\omega_\beta + \frac{\chi_1^2}{\Delta} \right) \\
\hline
E_{0,3} & 3\hbar \left(\omega_\beta + \frac{2\chi_1^2}{\Delta} \right) \\
\hline
\end{array}
\]

For a given $n_\alpha$, the Kerr term renders the spacing between the energy levels of the probe mode uneven as shown in the table. The spacing linearly increases with $n_\beta$ as
\begin{equation}\label{deltaE}
    \Delta_E(n_\beta)=E_{n_\alpha,n_\beta+1}-E_{n_\alpha,n_\beta}=\hbar\l[\omega_\beta-\frac{\chi_1^2}{\Delta}\l(4n_\alpha-2n_\beta\r)\r]
\end{equation}
Since the level-spacing is $n_\beta$-dependent, the resonant frequency of the probe mode for a given $n_\alpha$ is not fixed. %and may give rise to multiple peaks owing to multi-photon processes. 
The frequency-pulling term creates only a global energy shift in mode $\alpha$, which does not affect the spectroscopy.

\subsection{Simulation results}\label{subsec: Kerr-simulation}
\subsubsection{Zero temperature and strong driving}\label{subsubsec: Kerr-simulation-zeroT-strong}
%Figures [\ref{general-Kerr}] shows the spectroscopic plots when different terms are present in the Hamiltonian at $k_B T/\hbar \omega_\beta=0.01$ (i.e. $T\to 0$) and $d=0.01~ \hbar\omega_\beta$ (strong driving). 

Figures [\ref{lowT2}] shows the spectroscopic plots in presence of the Kerr term for different driving strengths at a very low temperature $k_B T/\hbar \omega_\beta=0.01$ (i.e. $T\to 0$).

A pair of well-resolved peaks appear for a given $n_\alpha$: one at $\omega_\text{res,1}^{n_\alpha}=\omega_\beta-4n_\alpha \chi_1^2/\Delta$ and the other at $\omega_\text{res,2}^{n_\alpha}=\omega_\beta-(4n_\alpha -1) \chi_1^2/\Delta$. The first peak arises due to the dispersive interaction and corresponds to one-photon absorption, whereas the second peak appears due to the Kerr term and corresponds to two-photon absorption \cite{Drummond_1980}. In the two-photon process, the bosonic mode absorbs two photons of frequency $\omega=\omega_\text{res,2}^{n_\alpha}$ for the transition $|n_\alpha\rangle|0\rangle\to|n_\alpha\rangle |2\rangle $, i.e. $E_{n_\alpha,2}-E_{n_\alpha,0}=2\hbar\omega_\text{res,2}^{n_\alpha}$. The difference in frequencies of one- and two-photon absorptions is a consequence of unequal level-spacings of the probe bosonic mode in presence of the Kerr term. In absence of the Kerr term, both processes occur at the same frequency $\omega=\omega_\text{res,1}^{n_\alpha}$. Since the two-photon transition probability goes as $\sim d^4$ while that of the single-photon transition as $\sim d^2$, the height of the peak at $\omega_\text{res,2}^{n_\alpha}$ increases more rapidly with the driving amplitude than the one at $\omega_\text{res,1}^{n_\alpha}$. 
    %\item The peak at $\omega_\text{res,2}^{n_\alpha}$ is narrower than the one at $\omega_\text{res,1}^{n_\alpha}$.
    
    Both sets of peaks fit the distribution $\{P_{n_\alpha}\}$ of the magnon mode but with different proportionality constants. Hence, the probe excitation at a resonance can be written as $\langle \beta^\dagger \beta\rangle_{n_\alpha,1(2)}=k_{1(2)} P_{n_\alpha}$.  This indicates that the magnon number states can still be resolved well at very low temperatures ($k_BT\ll \hbar\omega_\beta$) despite the presence of a Kerr-nonlinearity.
   % \item In absence of the dispersive interaction, the peak corresponding to $n_\alpha=0$ (which is equivalent to vanishing of the dispersive term) remains while the other peaks disappear. This means that the Kerr-like term produces the fundamental resonance at $\omega_\beta$ as well as one at a shifted frequency.  
    %\item The new peak is quite prominent.
    %\item A highly diminished peak also appears at $\omega_\beta-(4n_\alpha -2) \chi_1^2/\Delta$.

\subsubsection{High temperature and weak driving}\label{subsubsec: Kerr-simulation-highT-weak}

Figure [\ref{diff-T-d001}] shows the spectroscopic plots in presence of the Kerr term at different temperatures and a weak drive ($d/\hbar \omega_\beta =0.001$). For $T\neq0$, small peaks appear at $\omega_\beta-(4n_\alpha -2) \chi_1^2/\Delta$. This frequency corresponds to the energy difference $E_{n_\alpha,2}-E_{n_\alpha,1}$, which implies resonant transitions $|n_\alpha,1\rangle\to |n_\alpha,2\rangle$. This is caused by thermal occupation of $|n_\alpha,1\rangle$ at finite temperature. 

At higher temperature, higher energy levels of the bosonic mode are also significantly occupied and transitions $|n_\alpha,n_\beta\rangle\to |n_\alpha,n_\beta+1\rangle$ can occur. Hence, for a given $n_\alpha$, resonant peaks should appear at all $\omega_\text{res}^{n_\alpha,n_\beta}=\omega_\beta-(4n_\alpha -2n_\beta) \chi_1^2/\Delta$ [see Eq. (\ref{deltaE})], $n_\beta=0,1,2,\dots
$ being the thermally occupied levels from which the transitions take place. In absence of the Kerr term, only the peak at $\omega_\text{res}^{n_\alpha,0}(\equiv\omega_\text{res}^{n_\alpha})$ is present, for a given $n_\alpha$, at any temperature. Thus, the Kerr term creates additional probe frequencies at finite temperature. Since $\omega_\text{res}^{n_\alpha+j,2j}=\omega_\text{res}^{n_\alpha,0}$ for $j=1,2,\dots$, the peaks corresponding to them coincide. This changes the proportionality of the peak at $\omega_\text{res}^{n_\alpha,0}$ with the probability distribution $\l\{P_{n_\alpha}\r\}$.
Hence, the magnon number states cannot be resolved well at high temperature in the presence of a Kerr-nonlinearity.

\subsubsection{High temperature and strong driving}\label{subsubsec: Kerr-simulation-highT-strong}

\begin{figure}[h]
		\centering
\includegraphics[trim={0cm 0cm 0cm 0cm},clip,width=8cm]{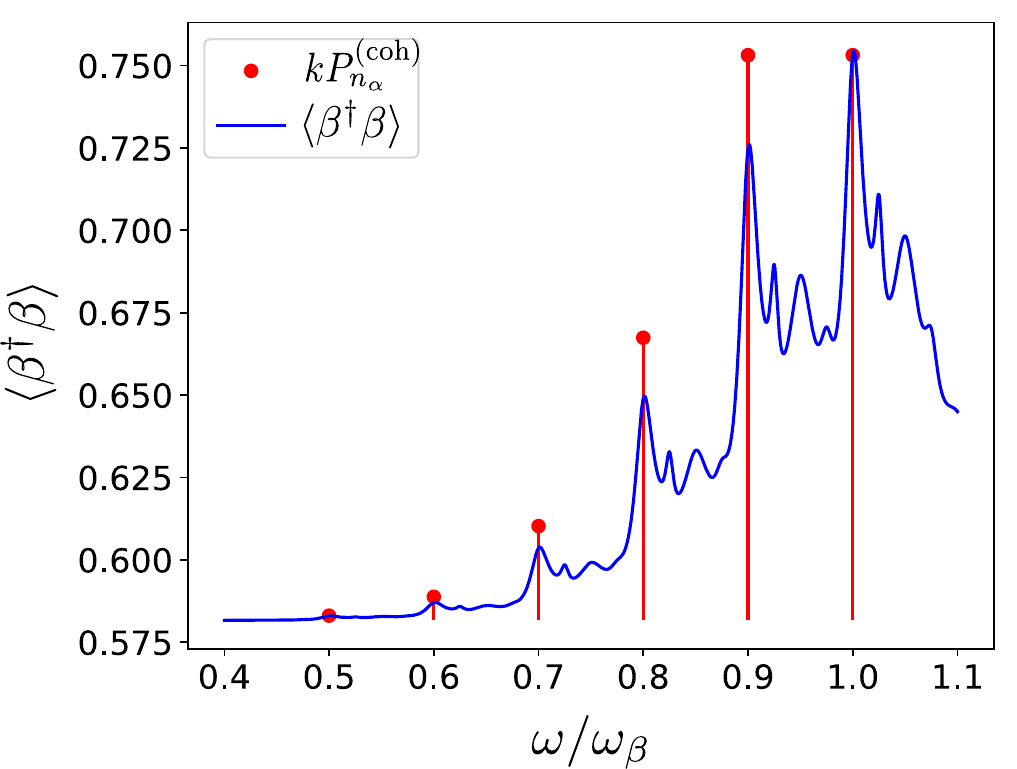}
		\caption{Simulation data for the probe bosonic mode at high temperature ($k_BT/\hbar \omega_\beta=1$) and strong driving ($d/\hbar\omega_\beta=0.01$) in presence of a Kerr nonlinearity. The magnon number states cannot be resolved because of the overlap of additional peaks resulting from  (i) two-photon absorptions at stronger driving and (ii) $|n_\alpha\rangle |n_\beta\rangle\to |n_\alpha\rangle |n_\beta+1\rangle$ transitions happening due to significant thermal occupation of excited levels $|n_\beta\rangle$ of the probe at high temperatures.}
		\label{highThighd}
	\end{figure}

Figure [\ref{highThighd}] shows the spectroscopic plots for higher temperature ($k_B T/\hbar \omega_\beta=1$)and stronger driving ($d=0.01~\hbar \omega_\beta$). In this case, the peaks due to both  thermal occupancy and two-photon absorptions appear. The peaks strongly overlap with each other and the magnon number states cannot be resolved. Thus, the bosonic mode cannot work as a sensor at high temperatures and strong drives in presence of a Kerr-nonlinearity.

\section{Proportionality of the spectroscopic peaks to magnonic probability distribution}\label{sec: proportionality}

\subsection{Time-dependent perturbation theory approach}\label{subsec: time-dependent}

In this section, we present a simple derivation using time-dependent perturbation theory to qualitatively show that the heights of the resonant peaks are proportional to the magnonic probability distribution. Let the system be initialized in a state $|\psi\rangle_\alpha|\xi\rangle_\beta$ where
\begin{equation}
|\psi\rangle_\alpha=\sum_{n^\prime_\alpha} c^\psi_{n^\prime_\alpha} |n^\prime_\alpha \rangle,~~~~~|\xi\rangle_\beta=\sum_{n^\prime_\beta} c^\xi_{n^\prime_\beta} |n^\prime_\beta \rangle.
\end{equation}
The magnonic probability distribution is $\{P_{n_\alpha^\prime}^{\psi}\}=\{|c_{n_\alpha^\prime}^{\psi}|^2\}$.
Using first-order time-dependent perturbation theory, the probability of the system being in state $|m_\alpha\rangle|n_\beta\rangle$ at time $t$ is
\begin{equation}
    P_{m_\alpha,n_\beta}(t)=\l|\langle m_\alpha|\langle n_\beta|\bigg(1-\frac{i}{\hbar}\int_0^t V_I(t^\prime) dt^\prime\bigg)|\psi\rangle_\alpha|\xi\rangle_\beta\r|^2=\l|c_{m_\alpha}^\psi c_{n_\beta}^\xi-\frac{i}{\hbar}\int_0^t\langle m_\alpha|\langle n_\beta| V_I(t^\prime) |\psi\rangle_\alpha|\xi\rangle_\beta dt^\prime\r|^2
\end{equation}
where
\begin{equation}
    V_I(t)=\e^{i\mathcal{H}_0 t/\hbar} V(t)\e^{-i\mathcal{H}_0 t/\hbar}=\e^{i\mathcal{H}_0 t/\hbar} [d\cos\omega t(\beta+\beta^\dagger) ]\e^{-i\mathcal{H}_0 t/\hbar}
\end{equation}
is the driving potential in the interaction picture with $\mathcal{H}_0=\hbar \omega_\alpha \alpha^\dagger \alpha + \hbar \omega_\beta \beta^\dagger \beta +\hbar \chi\alpha^\dagger \alpha \beta^\dagger \beta$. Discarding the counter-rotating terms, we get
\begin{equation}
P_{m_\alpha,n_\beta}(t)=|c_{m_\alpha}^\psi|^2\l|c_{n_\beta}^\xi-\frac{i ~d ~f_{n_\beta}}{\hbar}\int_0^t\cos[\omega-(\omega_\beta+m_\alpha \chi)]t^\prime dt^\prime\r|^2
\end{equation}

\begin{equation}
    P_{m_\alpha,n_\beta}(t)=P_{m_\alpha}^\psi\l|c_{n_\beta}^\xi-\l(\frac{i ~d~f_{n_\beta}}{\hbar}\r)\frac{\sin[\omega-(\omega_\beta+m_\alpha \chi)]t}{\omega-(\omega_\beta+m_\alpha \chi)} \r|^2
\end{equation}
where $f_{n_\beta}=c^\xi_{n_\beta+1}\sqrt{n_\beta +1}+c^\xi_{n_\beta-1}\sqrt{n_\beta}$. So, the mean probe excitation is
\begin{equation}
    \langle\beta^\dagger \beta\rangle=\sum_{m_\alpha,n_\beta} n_\beta P_{m_\alpha,n_\beta}(t)=\sum_{m_\alpha}P_{m_\alpha}^\psi\l(\sum_{n_\beta}n_\beta  \l|c_{n_\beta}^\xi-\l(\frac{i ~d~f_{n_\beta}}{\hbar}\r)\frac{\sin[\omega-(\omega_\beta+m_\alpha \chi)]t}{\omega-(\omega_\beta+m_\alpha \chi)} \r|^2\r)
\end{equation}
At very large times, the sum within the parenthesis peaks sharply at $\omega=\omega_\beta+m_\alpha \chi$. This explains the spectroscopic peaks qualitatively. For example, if the probe be initialized in vacuum i.e. $|\xi\rangle_\beta=|0\rangle$, then at $t\to\infty$,
\begin{equation}
    \langle\beta^\dagger \beta\rangle\approx\sum_{m_\alpha}\l(\frac{\pi d^2 }{\hbar^2 }\r)P_{m_\alpha}^\psi t ~\delta(\omega-\omega_\beta-m_\alpha \chi).
\end{equation}
which follows from $\lim_{t \to \infty}
 \frac{\sin^2 t x }{t x^2}=\pi \delta(x)$. %The proportionality constant is independent of the value of $m_\alpha$, due to which the probability distribution of the magnon mode can be resolved.

 \subsection{Lindblad master equation approach}\label{lindblad-section}
In this section, we provide a analytical derivation of the steady state spectrum of the probe bosonic mode using Lindblad master equation, which treats the probe as an open quantum system \cite{Breuer}.

 The Hamiltonian of a coherently driven probe bosonic mode ($\beta$) dispersively coupled to the magnon mode ($\alpha$) is expressed as
\begin{equation}
    H(t) = \hbar \omega_\alpha \alpha^\dagger \alpha + \hbar \omega_\beta \beta^\dagger \beta + \hbar \chi \alpha^\dagger \alpha \beta^\dagger \beta +  d \cos\omega t (\beta + \beta^\dagger)
\end{equation}
where $\chi$ signifies the strength of dispersive coupling and $d$ the driving strength.

We consider the probe mode to be coupled to a finite temperature bath. The Lindblad master equation in the Schr\"odinger picture is 
\begin{equation}\label{Lindblad}
    \frac{d \rho}{dt} = -\frac{i}{\hbar}[H(t),\rho] + \kappa(n_\text{th}^\beta+1) \mathcal{L}_\beta [\rho] +  \kappa~ n_\text{th}^\beta \mathcal{L_{\beta^\dagger}} [\rho]
\end{equation}
where 
\begin{equation}
    \mathcal{L}_a[b]= a b a^\dagger -\frac{1}{2}\l\{a^\dagger a, b\r\}
\end{equation}
is the dissipator. The jump operators are $\beta$ and $\beta^\dagger$, $\kappa$ is the probe decay rate and $n_\text{th}^\beta =[e^{\hbar \omega_\beta/k_B T}-1]^{-1}$ is the mean thermal occupation of the bath mode with frequency $\omega_\beta$.

Moving to a rotating frame ($U=\e^{-i\omega t \beta^\dagger \beta}$) and using the rotating wave approximation, we obtain a new Lindblad master equation,
\begin{equation}
    \frac{d \rho_r}{dt} = -\frac{i}{\hbar}[H_r,\rho_r] + \kappa(n_\text{th}^\beta+1) \mathcal{L}_\beta [\rho_r] +  \kappa~n_\text{th}^\beta \mathcal{L_{\beta^\dagger}} [\rho_r]
\end{equation}
where $\rho_r=U^\dagger \rho ~U$ and
\begin{equation}
    H_r=U^\dagger H U +i\hbar\l(\frac{dU^\dagger}{dt}\r)U=\hbar \omega_\alpha \alpha^\dagger \alpha+\hbar (\omega_\beta + \chi \alpha^\dagger \alpha-\omega)\beta^\dagger \beta   + \frac{d}{2} (\beta^\dagger+\beta)
\end{equation}
is time-independent. From the master equation, one can derive the following equation for the expectation value of an operator $O$:
\begin{equation}\label{trace}
\frac{d}{dt}\bigg{\langle} O\bigg{\rangle} = -\frac{i}{\hbar}\bigg{\langle} [O,H_r]\bigg{\rangle}+\kappa (n_\text{th}^\beta+1) \bigg{\langle} \bar{\mathcal{L}}_\beta[O]\bigg{\rangle} + \kappa ~n_\text{th}^\beta \bigg{\langle} \bar{\mathcal{L}}_{\beta^\dagger}[O] \bigg{\rangle} 
\end{equation}
where $\bigg{\langle}A \bigg{\rangle} = \text{Tr}~ (\rho_r A)$ and $\bar{\mathcal{L}}$ is the adjoint dissipator defined as
\begin{equation}
    \bar{\mathcal{L}}_a[b]= a^\dagger b a -\frac{1}{2}\l\{a^\dagger a, b\r\}.
\end{equation}
{\bf Case I: }Consider $O=\beta^\dagger \beta$. Using Eq. (\ref{trace}), we get
\begin{equation}
    \frac{d}{dt}\bigg{\langle}\beta^\dagger \beta\bigg{\rangle}=-\frac{id}{2\hbar}\l(-\bigg{\langle} \beta\bigg{\rangle}+\bigg{\langle} \beta^\dagger \bigg{\rangle}\r) - \kappa \l(\bigg{\langle}\beta^\dagger\beta\bigg{\rangle}-n_\text{th}^\beta\r).
\end{equation}
At steady state,
\begin{equation}\label{steady1}
    0=-\frac{id}{2\hbar}\l(-\bigg{\langle} \beta\bigg{\rangle}_\text{ss}+\bigg{\langle} \beta^\dagger \bigg{\rangle}_\text{ss}\r) - \kappa \l(\bigg{\langle}\beta^\dagger\beta\bigg{\rangle}_\text{ss}-n_\text{th}^\beta\r).
\end{equation}
\begin{equation}
    \implies 0=-\frac{id}{2\hbar}\l(-\sum_{n_\alpha}\langle n_\alpha|[ \text{Tr}_\beta (\rho_r \beta)]_\text{ss}|n_\alpha\rangle+\sum_{n_\alpha}\langle n_\alpha|[ \text{Tr}_\beta (\rho_r \beta^\dagger)]_\text{ss}|n_\alpha\rangle\r) - \kappa \l(\bigg{\langle}\beta^\dagger\beta\bigg{\rangle}_\text{ss}-n_\text{th}^\beta\r).
\end{equation}
where $\text{Tr}_\beta(A)=\sum_{n_\beta}\langle n_\beta|A|n_\beta\rangle$ is the reduced density matrix obtained by tracing out over $\beta$ subspace and subscript `ss' stands for steady-state value.

{\bf Case II: } Consider $O=\beta |n_\alpha\rangle\langle n_\alpha|$. Then, using Eq.(\ref{trace}), we get
\begin{equation}
    \frac{d}{dt}\bigg{\langle} \beta | n_\alpha\rangle \langle n_\alpha|  \bigg{\rangle}=-\frac{id}{2\hbar}\bigg{\langle}| n_\alpha\rangle \langle n_\alpha|\bigg{\rangle}-\frac{\kappa}{2}\bigg{\langle} \beta | n_\alpha\rangle \langle n_\alpha| \bigg{\rangle} -i(\omega_\beta-\omega)\bigg{\langle} \beta | n_\alpha\rangle \langle n_\alpha| \bigg{\rangle}-i\chi \bigg{\langle} n_\alpha \beta | n_\alpha\rangle \langle n_\alpha| \bigg{\rangle}.
\end{equation}
At steady state,
\begin{equation}\label{steady2}
    0 = -\frac{id}{2\hbar} \langle n_\alpha|[ \text{Tr}_\beta (\rho_r)]_\text{ss}|n_\alpha\rangle-\l[ \frac{\kappa}{2}+i(\omega_\beta+n_\alpha\chi-\omega)\r] \langle n_\alpha|[ \text{Tr}_\beta (\rho_r\beta)]_\text{ss}|n_\alpha\rangle.
\end{equation}
{\bf Case III: } Consider $O=|n_\alpha\rangle\langle n_\alpha|$. Using Eq.(\ref{trace}), we get 
\begin{equation}
\begin{aligned}
    &\frac{d}{dt}\bigg{\langle}  | n_\alpha\rangle \langle n_\alpha|  \bigg{\rangle} = 0\\
    &\implies \bigg{\langle}  | n_\alpha\rangle \langle n_\alpha|  \bigg{\rangle}=\text{const.}
    \\&\implies \sum_{n_\alpha^\prime}\langle n_\alpha^\prime | \text{Tr}_\beta \l(\rho_r\r) | n_\alpha\rangle \langle n_\alpha|n_\alpha^\prime\rangle=\text{const.}\\
    &\implies \langle n_\alpha| \text{Tr}_\beta \l(\rho_r\r) | n_\alpha\rangle =\text{const.}
    \end{aligned}
\end{equation}
\begin{equation}\label{prob}
    \implies \langle n_\alpha|[ \text{Tr}_\beta (\rho_r)]_\text{ss}|n_\alpha\rangle=  \langle n_\alpha|[ \text{Tr}_\beta (\rho_r)]_{t=0}|n_\alpha\rangle= \langle n_\alpha|[ \text{Tr}_\beta (\rho_\alpha\otimes\rho_\beta)]|n_\alpha\rangle = \langle n_\alpha| \rho_\alpha|n_\alpha\rangle= \rho_\alpha^{n_\alpha}=P_{n_\alpha}.
\end{equation}
The above result shows that the driving does not change the magnon number statistics $\{P_{n_\alpha}\}$, implying that it is a {\it Quantum Non-Demolition measurement}.
\\

Now, we combine the results of the three cases. Using Eq. (\ref{prob}) in Eq.(\ref{steady2}), we get
\begin{equation}\label{steady3}
    \langle n_\alpha|[ \text{Tr}_\beta (\rho_r\beta)]_\text{ss}|n_\alpha\rangle=\frac{-i P_{n_\alpha} d }{2\hbar\l[\frac{\kappa}{2}+i\l(\omega_\beta+\chi n_\alpha - \omega\r)\r]}
\end{equation}
Furthermore, $\langle n_\alpha|[ \text{Tr}_\beta (\rho_r\beta^\dagger)]_\text{ss}|n_\alpha\rangle=\langle n_\alpha|[ \text{Tr}_\beta (\rho_r\beta)]_\text{ss}|n_\alpha\rangle^*$.
Using Eq.(\ref{steady3}) and its conjugate in Eq. (\ref{steady1}), we get
\begin{equation}
    0=-\l(\frac{id}{2\hbar}\r)\sum_{n_\alpha}\l(\frac{iP_{n_\alpha}d }{2\hbar}\r)\l[ \frac{1}{\frac{\kappa}{2}+i\l(\omega_\beta+\chi n_\alpha - \omega\r)}+\frac{1}{\frac{\kappa}{2}-i\l(\omega_\beta+\chi n_\alpha - \omega\r)}\r]- \kappa \l(\bigg{\langle}\beta^\dagger\beta\bigg{\rangle}_\text{ss}-n_\text{th}^\beta\r)
\end{equation}
\begin{equation}
    \implies \bigg{\langle}\beta^\dagger\beta\bigg{\rangle}_\text{ss}=\sum_{n_\alpha}P_{n_\alpha}\l(\frac{(d/\hbar)^2}{\kappa^2+4[(\omega_\beta+\chi n_\alpha)-\omega]^2}\r)+n_\text{th}^\beta.
\end{equation}
Hence, the steady state excitation of the bosonic mode has a series of resonances at $\omega=\omega_\beta+\chi n_\alpha$ and the heights of the resonant peaks are proportional to $P_{n_\alpha}$. Each resonant peak has an identical constant of proportionality $d^2/\hbar^2\kappa^2$ which enables the sensing of the magnonic probability distribution.

\section{Design Principles}\label{sec: Design}
In this section, we provide the design principles to engineer a strong dispersive interaction between the magnon mode and the probe bosonic mode. Although the design principles are applicable to any bosonic mode suited to the specific platform, we cite two candidate materials -- hematite nanoparticle and Yttrium iron garnet (YIG) thin film -- that might be useful to perform the sensing. 

\subsection{Antiferromagnetic magnon mode as probe}\label{subsec: afm mode}

The strength of the direct dispersive coupling between the two magnon modes of an AFM would be $\chi = \dfrac{Jq}{\hbar N}$, as shown in Eq.(\ref{dispersive}). The value of exchange interaction $J$ is usually given in the literature in terms of the exchange field $\mu_0 H_E$. In the following, we deduce a relation between them. The Heisenberg exchange Hamiltonian of an AFM is expressed as
\begin{equation}
\begin{aligned}
   \mathcal{H}_\text{E}&=\frac{J}{\hbar^2}\sum_{i}^{}\sum_{\boldsymbol{\delta}}^{}{\bf S}_\text{A}({\bf r}_i)\cdot {\bf S}_\text{B}({\bf r}_i+\boldsymbol{\delta})\\
   &\approx\frac{Jq}{\hbar^2}\sum_{i}^{}{\bf S}_\text{A}({\bf r}_i)\cdot {\bf S}_\text{B}({\bf r}_i)\\
   &=\frac{Jq}{\gamma^2\hbar^2}\sum_{i}^{}\boldsymbol{\mu} _\text{A}({\bf r}_i)\cdot \boldsymbol{\mu}_\text{B}({\bf r}_i)\\
   &= \frac{Jq}{\gamma^2\hbar^2 (\Delta V)}\int\boldsymbol{\mu} _\text{A}({\bf r})\cdot \boldsymbol{\mu}_\text{B}({\bf r})d^3{\bf r}\\
   &= \frac{Jq M_s^2 (\Delta V)^2}{\gamma^2\hbar^2 (\Delta V)}\int \hat{\bf m} _\text{A}({\bf r})\cdot \hat{\bf m}_\text{B}({\bf r})d^3{\bf r}\\
   &=
   \frac{Jq M_s^2 (\Delta V)}{\gamma^2\hbar^2}\int \hat{\bf m} _\text{A}({\bf r})\cdot \hat{\bf m}_\text{B}({\bf r})d^3{\bf r}\\
   &=(\mu_0 H_E) M_s\int \hat{\bf m} _\text{A}({\bf r})\cdot \hat{\bf m}_\text{B}({\bf r})d^3{\bf r}
   \end{aligned}
\end{equation}
where $\boldsymbol{\mu}_{A,B}$ are the magnetic moments at a lattice site, $q$ is the coordination number, $\mu_0 H_E = \dfrac{JqM_s (\Delta V)}{\gamma^2\hbar^2}$ is the exchange field, $\Delta V$ is the volume of the unit cell and $M_s$ is the magnetization of each sublattice. This gives $J=\dfrac{\mu_0 H_E\gamma^2\hbar^2}{q M_s (\Delta V)}$.
Hence, the dispersive strength is 
\begin{equation}\label{design-magnon}
    \chi = \dfrac{\mu_0 H_E\gamma^2\hbar}{ M_s V},
\end{equation}
where $V=N(\Delta V)$ is the total volume of the sample. Equation (\ref{design-magnon}) provides the design guidance to use an AFM magnon mode as the probe. %The coupling strength is an extrinsic quantity as it depends on the volume of the sample. 
Since the dispersive strength is inversely proportional to the volume and directly proportional to the exchange field, it can be enhanced by choosing smaller AFMs with higher exchange fields (e.g. hematite where $\mu_0H_E\approx1000 $ T.).

Now, we estimate the strength of the dispersive coupling in hematite. Hematite is a canted AFM at room temperature that shows a weak spontaneous magnetization. The weak magnetization ($m_s$) is related to sublattice magnetization ($M_s$) as $m_s=2M_s\l(\dfrac{H_D+H}{2H_E+H_a}\r)$, where $H_E$, $H_D$,  $H_a$ and $H$ denote the magnetic field strengths corresponding to the antiferromagnetic exchange interaction, Dzyaloshinskii-Moriya interaction, easy-axis anisotropy in the basal plane, and the applied field respectively \cite{Pincus1960}. Using values of the parameters given in Ref. \cite{Boventer2021} -- $m_s \approx 3\times10^3$ A/m, $\mu_0 H_E=1000$ T, $\mu_0 H_D = 2.26$ T, $\mu_0 H_a=6\times10^{-5} $ T, $\mu_0 H=0.2$ T, we get $M_s\approx1.22\times10^{6}$ A/m. We estimate that the dispersive shift $\dfrac{\chi}{2\pi}$ can be of the order of a sub- or a few GHz for hematite nanoparticles.
%\begin{center}
    %{\color{magenta}For a spherical hematite nanoparticle of radius 3 nm, the dispersive shift in frequency is obtained as $\dfrac{\chi}{2\pi}\approx3.77$ GHz.}
%\end{center}
The frequency of the quasi-ferromagnetic resonance mode of hematite, which is to be used as the probe magnon mode, at $\mu_0 H=0.2$ T is $f\approx22 $ GHz. If the sample quality factor is $Q=500$ \cite{Boventer2021}, the frequency linewidth is $\Delta f = 44$ MHz. So, the resolution of the spectroscopic peaks should be possible for hematite nanoparticles. 

\subsection{Phonon mode as probe}\label{subsec: phonon mode}

In this section, we derive the expression for a nonlinear magnon-phonon coupling in a ferromagnet for a specific setup, which gives the design guidance to use a phonon mode as the probe for sensing the magnon number states.

\begin{figure}[htbp]
		\centering
\includegraphics[trim={0cm 0cm 0cm 0cm},clip,width=12cm]{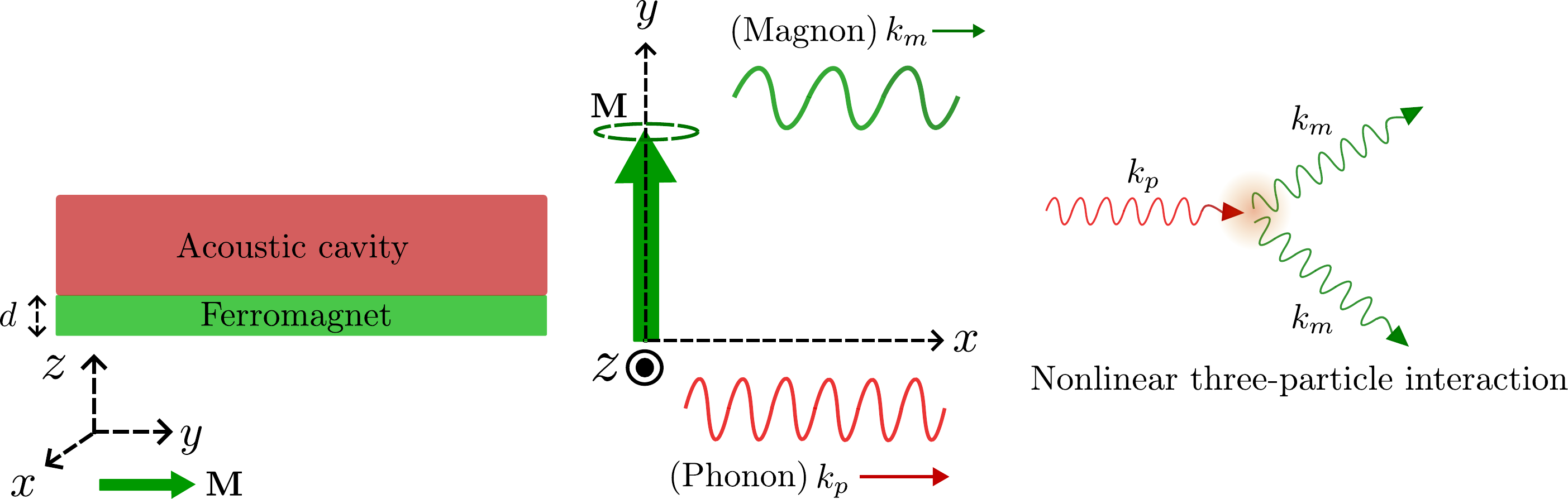}
		\caption{Schematic of a hybrid heterostructure of an acoustic cavity and a FM thin film which may be used to realize a nonlinear magnon-phonon coupling and utilize the phonon mode as the probe.}
		\label{acoustic}
	\end{figure}

Consider a planar heterostructure of an acoustic cavity fabricated on a ferromagnetic (FM) thin film. In recent times, these systems have been extensively studied to explore the coupling between surface acoustic waves (SAWs) and spin waves \cite{kuenstle2025,Hwang_Otani_2025}. The SAWs are excited in the acoustic cavity and they interact with the magnon modes of the FM film. This phenomenon is captured by the magnetoelastic coupling,
\begin{equation}\label{Hme}
    \mathcal{H}_\text{me} = \frac{1}{M_s^2}\int_\text{FM} d^3{\bf r}\left[ b_1\left(m_x^2 e_{xx} + m_y^2 e_{yy}+ m_z^2 e_{zz}\right)+ 2 b_2\l(m_x m_y e_{xy}+m_y m_z e_{yz} + m_x m_z e_{xz}\r)\right]
\end{equation}
where $b_{1,2}$ are the magnetoelastic coefficients of the FM, $m_i$ are the components of the magnetization and $e_{ij}$ is the strain tensor with $i=j$ and $i\neq j$ denoting longitudinal and shear strains respectively. The magnetization components and the strain tensors in Eq. (\ref{Hme}) are quantized to derive the required three-particle magnon-phonon interactions \cite{kittel1963quantum}.

Let the surface of the FM film be parallel to $x$-$y$ plane [Fig. \ref{acoustic}]. The saturation magnetization is in-plane, say ${\bf M}=M_s \hat{y} $. We consider a special case where only a longitudinal phonon mode with wave vector ${\bf k}=k_p \hat{x}$ is excited in the cavity. We further assume the thin-film limit so that the displacement or magnetization components have negligible spatial variation along the thickness of the film i.e. along $z$ direction. Then, the quantized displacement operator associated with the phonon mode can be expressed as
\begin{equation}u_x=\sqrt{\frac{\hbar}{\rho V \omega_p}}\sin(k_p x) (\beta+\beta^\dagger),
\end{equation}
where $\rho$ is the mass density of the FM, $V$ is the volume of the FM film, $\omega_\beta$ is the phonon frequency,  $\beta$ and $\beta^\dagger$ are the phonon annihilation and creation operators, respectively. Similarly, the transverse component of the magnetization (along the direction of the acoustic displacement) is expressed as
\begin{equation}m_x=\sqrt{\frac{\gamma M_s \hbar}{V}}\cos (k_m x) \l(\alpha+\alpha^\dagger\r),
\end{equation}
where $k_m$ is the wave number of the magnon mode and $\alpha$ and $\alpha^\dagger$ are its annihilation and creation operators, respectively.
Since we consider only a longitudinal phonon mode along the $x$ direction, only the strain $e_{xx}$ comes into play in the magnetoelastic coupling. Here, 
\begin{equation}
    e_{xx}= \frac{\partial u_x}{\partial x} = \sqrt{\frac{\hbar}{\rho V \omega_\beta}} k_p\cos(k_p x) (\beta+\beta^\dagger).
\end{equation}
Hence, the magnetoelastic coupling for this setup reduces to
\begin{equation}\label{nonlinear}
\begin{aligned}
    H_\text{me}&= \frac{b_1}{M_s^2}\int_\text{FM}m_x^2 e_{xx} d^3{\bf r}\\
    &=\l(k_p\frac{b_1\gamma \hbar L_y d }{M_s  V}\r)\sqrt{\frac{\hbar}{\rho V \omega_\beta}}\mathcal{I}\l[\l(\alpha+\alpha^\dagger\r)^2\l(\beta+\beta^\dagger\r)\r]\\&=\l(k_p\frac{b_1\gamma \hbar L_y d }{M_s  V}\r)\sqrt{\frac{\hbar}{\rho V \omega_\beta}}\mathcal{I}\l[\alpha^{\dagger2}\beta + \alpha^{\dagger2} \beta^\dagger +\l(2\alpha^\dagger \alpha+1\r)\beta + \text{h.c.}\r]
\end{aligned}
\end{equation}
where $\mathcal{I}=\int_{-\frac{L_x}{2}}^{\frac{L_x}{2}}\cos^2(k_m x)\cos(k_p x) dx$.
As discussed in Sec. \ref{effective-dispersive}, the coupling $\hbar \chi_3 (\alpha^{\dagger2}\beta + \text{h.c.})$ can give rise to an effective dispersive interaction $\hbar\chi_\text{eff}\alpha^\dagger \alpha \beta^\dagger \beta$ in far detuned limit $(\chi_3 \ll |\omega_\beta-2\omega_\alpha|)$. We focus only on this coupling term in Eq. (\ref{nonlinear}), which corresponds to the conversion of a phonon into two magnons (parametric down-conversion). This requires $2k_m=k_p$ by momentum conservation.
The strain $e_{xx}$ should vanish at the boundary, which requires $\cos\l(\frac{ k_p L_x}{2}\r)=0$. Using this boundary condition and assuming that the wavelength of the phonon mode is much smaller than the length of the sample \cite{Hwang_Otani_2025}, we get $\mathcal{I}\approx L_x/4$. Then, we get the desired nonlinear coupling strength as
\begin{equation}
    \chi_3=\l(\frac{\pi b_1\gamma  }{2M_s  \lambda_p}\r)\sqrt{\frac{\hbar}{\rho V \omega_\beta}},
\end{equation}
where $\lambda_p=2\pi/k_p$ is the wavelength of the phonon mode. The \textit{effective} dispersive shift in the frequency is $\chi_\text{eff}/2\pi$ where 
\begin{equation}\label{phonon-design}
    \chi_\text{eff}=\frac{4\chi_3^2}{\omega_\beta-2\omega_\alpha}=\l(\frac{1}{\omega_\beta-2\omega_\alpha}\r)\l(\frac{\pi b_1\gamma  }{M_s  \lambda_p}\r)^2\l(\frac{\hbar}{\rho V \omega_\beta}\r).
\end{equation}
Equation (\ref{phonon-design}) provides the general design principles to achieve a dispersive shift which can be sufficiently larger than the linewidth of phonon resonance. The factor $1/(\omega_\beta-2\omega_\alpha)$ in Eq.(\ref{phonon-design})  offers tunability  as the frequencies can be optimally chosen to enhance $\chi_\text{eff}$ under the constraint $\chi_3\ll\omega_\beta-2\omega_\alpha$. Since $2\omega_\alpha \lesssim \omega_\beta$, this method is more suitable to resolve the number states of a low frequency magnon mode.

Now, we estimate the effective dispersive strength (\ref{phonon-design}) if YIG is used as the material for the ferromagnetic film. The relevant parameters for it are:  $b_1=3.48\times 10^{5}~\text{Jm}^{-3}, ~ M_s=127\times10^3\text{A/m},~\rho=5110\text{ Kg m}^{-3}$. For a phonon mode of frequency $\omega_\beta=2\pi\times 1$ GHz, wavelength $\lambda_p=10 $ nm and a 100 nm $\times$ 100 nm $\times$ 1 nm thin YIG film, we get the coupling strength $\chi_3\approx43$ MHz. Considering a modest detuned regime such that $\chi_3/(\omega_\beta-2\omega_\alpha)=1/6$, the dispersive shift $\chi_\text{eff}/2\pi\approx2.61 $ MHz. It is feasible to achieve phonon linewidths of the order of a few MHz \cite{kuenstle2025}. The dispersive effects might still be observable using a phononic probe although the peaks may not be well-resolved \cite{Quirion2017}. Furthermore, as $\chi_\text{eff}\propto b_1^2$, materials with higher magnetoelastic coefficients can significantly enhance the effective dispersive coupling.

For comparative analysis, we tabulate the parameters used in different works where a qubit was used to probe the boson number states [see Table \ref{tab:qubit_mode_comparison}].

\begin{table}[htbp]
\centering
\caption{Comparison of qubit--bosonic mode coupling parameters (all frequencies given as $/2\pi$).}
\begin{tabular}{llcccccc}
\hline
Mode to  & Reference & Detuning  &  $~~~~$Linear coupling strength & $g/\Delta$ & Dispersive Shift  & Qubit decay rate \\
 be sensed&  & $\Delta$ (MHz) & $g$ (MHz) &  & $2\chi=2g^2/\Delta$ (MHz) & (MHz) \\
\hline
Phonon & \cite{Arrangoiz-Arriola2019}   & 88    & 15.2  & $1/6$    & 3.00  & 0.60 \\
Phonon & \cite{Lee2023}       & ---   & 13.5  & $1/9$    & 2.23  & 0.25 \\
Photon & \cite{Schuster2007}  & 1200  & 105   & $1/11.5$ & 17.0  & 1.80 \\
Magnon & \cite{Quirion2017}   & 89    & 7.79  & $1/11.4$ & 1.50  & 0.78 \\
\hline
\end{tabular}
\label{tab:qubit_mode_comparison}
\end{table}

\end{document}